\RequirePackage{etoolbox}
\csdef{input@path}{{style/}{graphics/}}
\documentclass[aop,MSNbibl,seceqn,dvips]{arximspdf}
\usepackage{mathbh}
\usepackage{graphicx}

\doi{10.1214/13-AOP843} %kopijuoti is PTS
\volume{43}
\issue{2}
\pubyear{2015}
\firstpage{639}
\lastpage{681}

\makeatletter
\mathchardef\varPi="0105
\renewcommand{\&}{and}
\newcommand{\odd}{\ \mathrm{odd}}
\newcommand{\even}{\ \mathrm{even}}
\newcommand{\underset}[2]{\mathop{#2}_{#1}}
\newcommand{\undersett}[2]{\mathop{#2}\limits_{#1}}
\newtheorem{theorem}{Theorem}[section]
\newproclaim{assumption}[theorem]{Assumption}
\newtheorem{lemma}[theorem]{Lemma}
\newtheorem{proposition}[theorem]{Proposition}
\newproclaim{remark}[theorem]{Remark}
\newcommand{\veee}[1]{|\!|\!|#1|\!|\!|}
\newcommand{\ising}{\mathrm{Ising}}
\newcommand{\perc}{\mathrm{perc}}
\newcommand{\rw}{\mathrm{RW}}
\newcommand{\saw}{\mathrm{SAW}}
\makeatother

\begin{document}
\begin{frontmatter}

\title{Critical two-point functions for long-range
statistical-mechanical models in high dimensions}
\runtitle{Critical two-point functions for long-range models}

\begin{aug}
\author[A]{\fnms{Lung-Chi} \snm{Chen}\ead[label=e1]{lcchen@math.fju.edu.tw}\thanksref{T1}}
\and
\author[B]{\fnms{Akira} \snm{Sakai}\corref{}\ead[label=e2]{sakai@math.sci.hokudai.ac.jp}\thanksref{T2}}
\runauthor{L.-C. Chen and A. Sakai}
\affiliation{Fu-Jen Catholic University and Hokkaido University}
\address[A]{Department of Mathematics\\
Fu-Jen Catholic University\\
510 Chung Cheng Road\\
Hsinchuang\\
Taipei County 24205\\
Taiwan\\
\printead{e1}} %adresu isvedimo komanda gale!
\address[B]{Department of Mathematics\\
Hokkaido University\\
North 10, West 8, Kita-ku\\
Sapporo\\
Hokkaido 060-0810\\
Japan\\
\printead{e2}}
\end{aug}
\thankstext{T1}{Supported in part by NSC Grant 99-2115-M-030-004-MY3 and in
part by NSC Grant 102-2115-M-030-001-MY2.}
\thankstext{T2}{Supported in part by JSPS Grant-in-Aid for Young
Scientists (B) 21740059 and in part by JSPS Grant-in-Aid for Scientific Research (C) 24540106.}

% HISTORY:
\received{\smonth{10} \syear{2012}}
\revised{\smonth{3} \syear{2013}}

% ABSTRACT
%
\begin{abstract}
We consider long-range self-avoiding walk, percolation and the Ising model
on $\mathbb{Z}^d$ that are defined by power-law decaying pair
potentials of
the form
$D(x)\asymp|x|^{-d-\alpha}$ with $\alpha>0$. The upper-critical dimension
$d_{\mathrm{c}}$ is $2(\alpha\wedge2)$ for self-avoiding walk and
the Ising model, and
$3(\alpha\wedge2)$ for percolation. Let $\alpha\ne2$ and assume certain
heat-kernel bounds on the $n$-step distribution of the underlying
random walk.
We prove that, for $d>d_{\mathrm{c}}$ (and the spread-out parameter
sufficiently large),
the critical two-point function $G_{p_{\mathrm{c}}}(x)$ for each model
is asymptotically
$C|x|^{\alpha\wedge2-d}$, where the constant $C\in(0,\infty)$ is expressed
in terms of the model-dependent lace-expansion coefficients and exhibits
crossover between $\alpha<2$ and $\alpha>2$. We also provide a~class of
random walks that satisfy those heat-kernel bounds.
\end{abstract}

% KEYWORDS

\end{frontmatter}

\setcounter{footnote}{2}
%s1 #&#
\section{Introduction}
The two-point function is one of the key observables to understand phase
transitions and critical behavior. For example, the two-point function for
the Ising model indicates how likely the spins located at those two sites
point in the same direction. If it decays fast enough to be summable, then
there is no macroscopic order.
The summability of the two-point function is lost as soon as the model
parameter (e.g., temperature) is above the critical point and, therefore,
it is naturally hard to investigate critical behavior.

The lace expansion is a powerful tool to rigorously prove mean-field behavior
above the model-dependent critical dimension. The mean-field behavior here
is for the two-point function at the critical point to exhibit similar
behavior to the underlying random walk. It has been successful to prove
such behavior for various statistical-mechanical models, such as self-avoiding
walk, percolation, lattice trees/animals and the Ising model. The best
lace-expansion result obtained so far is to identify an asymptotic expression
($={}$the Newtonian potential times a model-dependent constant) of the critical
two-point function for finite-range models, such as the nearest-neighbor
model. However, this ultimate goal has not been achieved before this paper
for long-range models, especially when the 1-step distribution for the
underlying random walk decays in powers of distance; only the infrared bound
on the Fourier transform of the two-point function was available. This was
partly because of our poor understanding of the long-range models in the
$x$-space, not in the Fourier space. For example, the random-walk Green's
function is known to be asymptotically Newtonian/Riesz depending on the power
of the aforementioned power-law decaying 1-step distribution, but we were
unable to find optimal error estimates in the literature. Also, the
subcritical two-point function is known to decay exponentially for the
finite-range models, but this is not the case for the power-law decaying
long-range models; as is shown in this paper, the decay rate of the
subcritical two-point function is the same as the 1-step distribution
of the
underlying random walk.

Therefore, the goal of this paper is to overcome those difficulties and
derive an asymptotic expression of the critical two-point function for
the power-law decaying long-range models above the critical dimension,
using the lace expansion. We would also like to investigate crossover
in the asymptotic expression when the power of the 1-step distribution
of the underlying random walk changes.

%s1.1 #&#
\subsection{Models and known results}\label{ssmodels}
Self-avoiding walk (SAW) is a model for linear polymers. We define the
two-point function for SAW on $\mathbb{Z}^d$ as
%
%e1.1 #&#
\begin{equation}
\label{eq-SAW-2pt} G_p^\saw(x)=\sum
_{\omega\dvtx o\to x}p^{|\omega|}\prod_{j=1}^{|\omega|}
D(\omega_j-\omega_{j-1})\prod
_{s<t}(1-\delta_{\omega_s,\omega_t}),
\end{equation}
where $p\ge0$ is the fugacity, $|\omega|$ is the length of a path
$\omega=(\omega_0,\omega_1,\ldots,\omega_{|\omega|})$ and
$D\dvtx\mathbb{Z}^d\to
[0,1]$ is
the $\mathbb{Z}^d$-symmetric nondegenerate [i.e., $D(o)\ne1$] 1-step
distribution for
the underlying random walk (RW);
the contribution from the 0-step walk is considered to be $\delta_{o,x}$
by convention.
If the\vspace*{1pt} indicator function $\prod_{s<t}(1-\delta_{\omega_s,\omega
_t})$ is
replaced by 1, then $G_p^\saw(x)$ turns into the RW Green's function
$G_p^\rw(x)$, whose radius of convergence $p_{\mathrm{c}}^\rw$ is 1, as
$\chi_p^\rw\equiv\sum_{x\in\mathbb{Z}^d}G_p^\rw(x)=(1-p)^{-1}$
for $p<1$ and
$\chi_p^\rw=\infty$ for $p\ge1$. Therefore, the radius of convergence
$p_{\mathrm{c}}^\saw$ for $G_p^\saw(x)$ is not less than 1. It is
known that
$\chi_p^\saw\equiv\sum_{x\in\mathbb{Z}^d}G_p^\saw(x)<\infty$ if
and only if
$p<p_{\mathrm{c}}^\saw$ and diverges as $p\uparrow p_{\mathrm
{c}}^\saw$. Here, and in the
remainder of the paper, we often use ``$\equiv$'' for definition.

Percolation is a model for random media. Each bond $\{u,v\}$, which is
a pair
of vertices in $\mathbb{Z}^d$, is either occupied or vacant
independently of the other
bonds. The probability that $\{u,v\}$ is occupied is defined to be $pD(v-u)$,
where $p\ge0$ is the percolation parameter. Since $D$ is a probability
distribution, the expected number of occupied bonds per vertex equals
$p\sum_{x\ne o}D(x)=p(1-D(o))$. The percolation two-point function
$G_p^\perc(x)$ is defined to be the probability that there is a self-avoiding
path of occupied bonds from $o$ to $x$; again by convention, $G_p^\perc(o)=1$.

The Ising model is a model for magnets. For $\Lambda\subset\mathbb
{Z}^d$ and
$\varphi=\{\varphi_v\}_{v\in\Lambda}\in\{\pm1\}^\Lambda$, we
define the
Hamiltonian (under the free-boundary condition) as
%
%e1.2 #&#
\begin{equation}
H_\Lambda(\varphi)=-\sum_{\{u,v\}\subset\Lambda}J_{u,v}
\varphi_u \varphi_v,
\end{equation}
where $J_{u,v}=J_{o,v-u}\ge0$ is the ferromagnetic pair potential and inherits
the properties of the given $D$, as explained below. The finite-volume
two-point function at the inverse temperature $\beta\ge0$ is defined as
%
%e1.3 #&#
\begin{equation}
\label{eq-therm-av} \langle\varphi_o\varphi_x
\rangle_{\beta,\Lambda}=\sum_{\varphi
\in\{\pm1\}
^\Lambda}
\varphi_o\varphi_x e^{-\beta H_\Lambda(\varphi)} \biggm/\sum
_{\varphi\in
\{\pm
1\}^\Lambda}e^{-\beta H_\Lambda(\varphi)}.
\end{equation}
It is known that $\langle\varphi_o\varphi_x\rangle_{\beta,\Lambda
}$ is
increasing in
$\Lambda\uparrow\mathbb{Z}^d$. Let $p=\sum_{x\in\mathbb
{Z}^d}\tanh(\beta J_{o,x})$. The Ising
two-point function $G_p^\ising(x)$ is defined to be the increasing-volume
limit of $\langle\varphi_o\varphi_x\rangle_{\beta,\Lambda}$:
%
%e1.4 #&#
\begin{equation}
G_p^\ising(x)=\lim_{\Lambda\uparrow\mathbb{Z}^d}\langle\varphi
_o\varphi_x\rangle_{\beta,
\Lambda}.
\end{equation}
Let $D(x)=p^{-1}\tanh(\beta J_{o,x})$.

For percolation and the Ising model, there is a model-dependent critical
point $p_{\mathrm{c}}\ge1$ (from now on, we omit the superscript,
unless it
causes any
confusion) such that
%
%e1.5 #&#
\begin{eqnarray}
\chi_p&\equiv&\sum_{x\in\mathbb{Z}^d}G_p(x)
\cases{ <\infty, &\quad$[p<p_{\mathrm{c}}]$,
\cr
=\infty, &\quad$[p\ge
p_{\mathrm{c}}]$,}
\nonumber\\[-8pt]\\[-8pt]
\theta_p&\equiv&\sqrt{\lim_{|x|\to\infty}G_p(x)}
\cases{ =0,&\quad$[p<p_{\mathrm{c}}]$,
\cr
>0,&\quad$[p>p_{\mathrm{c}}]$.}\nonumber
\end{eqnarray}
The order parameter $\theta_p^\perc$ is the probability that the occupied
cluster of the origin~is unbounded, while $\theta_p^\ising$ is the spontaneous
magnetization, which is the infinite-volume limit of the finite-volume
single-spin expectation $\langle\varphi_o\rangle_{\beta,\Lambda
}^+$ under the
plus-boundary condition. The continuity of $\theta_p$ at $p=p_{\mathrm
{c}}$ in a general
setting is still a~remaining issue.

We are interested in asymptotic behavior of $G_{p_{\mathrm{c}}}(x)$ as
$|x|\to
\infty$.
For the ``uniformly spread-out'' finite-range models, for example,
$D(x)={\mathbh1}_{\{|x|=1\}}/\break (2d)$ or $D(x)={\mathbh1}_{\{\|x\|
_\infty\le L\}}/(2L+1)^d$ for some
$L\in[1,\infty)$, it has been proved \cite{h08,hhs03,s07} that, if
$d>4$ for
SAW and the Ising model and $d>6$ for percolation, and if $d$ or $L$ is
sufficiently large (depending on the models), then there is a model-dependent
constant $A$ ($=1$ for RW) such that
%
%e1.6 #&#
\begin{equation}
\label{eq-Gpc-fr} G_{p_{\mathrm{c}}}(x)\underset{|x|\to\infty}\sim\frac
{a_d/\sigma
^2}{A|x|^{d-2}},
\end{equation}
where ``$\sim$'' means that the asymptotic ratio of the left-hand side
to the
right-hand side is 1, and
%
%e1.7 #&#
\begin{equation}
\label{eq-variance} a_d=\frac{d\Gamma((d-2)/2)}{2\pi^{d/2}},\qquad \sigma^2
\equiv\sum_{x\in\mathbb{Z}^d}|x|^2D(x)=O
\bigl(L^2 \bigr).
\end{equation}
This is a sufficient condition for the following mean-field behavior
\cite{a82,af86,an84,ba91,ms93}:
%
%e1.8 #&#
\begin{eqnarray}
\label{eq-MF} \chi_p\underset{p\uparrow p_{\mathrm{c}}}
\asymp(p_{\mathrm{c}}-p)^{-1},\qquad \theta_p
\underset{p \downarrow p_{\mathrm{c}}}\asymp\cases{ \displaystyle
\sqrt{p-p_{\mathrm{c}}}, &\quad[Ising],
\cr
\displaystyle p-p_{\mathrm{c}},
&\quad[percolation],}
\end{eqnarray}
where ``$\asymp$'' means that the asymptotic ratio of the left-hand
side to
the right-hand side is bounded away from zero and infinity.

The proof of the above result is based on the lace expansion (e.g.,
\cite{hs90,ms93,s07,s06}). The core concept of the lace expansion is to
systematically isolate interaction among individuals (e.g., mutual avoidance
between distinct vertices for SAW or between distinct occupied pivotal bonds
for percolation) and derive macroscopic recursive structure that yields the
random-walk like behavior (\ref{eq-Gpc-fr}). When $d>d_{\mathrm{c}}$
and $d\vee L\gg1$
(i.e., $d$ or $L$ sufficiently large depending on the models), there is
enough room for those individuals to be away from each other, and the lace
expansion converges \cite{hs90,ms93,s07,s06}. The resultant recursion equation
for $G_p$ is the following:
%
%e1.9 #&#
\begin{eqnarray}
G_p(x)= \cases{ \displaystyle\delta_{o,x}+\sum
_{v\in\mathbb{Z}^d}pD(v) G_p(x-v),\qquad\mbox{[RW],}
\vspace*{6pt}\cr
\displaystyle\delta_{o,x}+\sum_{v\in\mathbb{Z}^d}
\bigl(pD(v)+\pi_p(v) \bigr) G_p(x-v),
\cr
\hspace*{156pt}\mbox{[SAW],}
\vspace*{6pt}\cr
\displaystyle\pi_p(x)+\mathop{\sum_{u,v\in\mathbb{Z}^d}}_{(u\ne
v)}
\pi_p(u) pD(v-u) G_p(x-v),
\cr
\hspace*{156pt} \mbox{[Ising \& percolation],}}\hspace*{-30pt}
\end{eqnarray}
where $\pi_p$ is the lace-expansion coefficient.
To treat all models simultaneously, we introduce the notation $f*g$
to denote the convolution of functions $f$ and $g$ in $\mathbb{Z}^d$:
%
%e1.10 #&#
\begin{equation}
(f*g) (x)=\sum_{v\in\mathbb{Z}^d}f(v) g(x-v).
\end{equation}
Then the above identities can be simplified as (the spatial variables are
omitted)
%
%e1.11 #&#
\begin{equation}
G_p= \cases{ \displaystyle\delta+pD*G_p, &\quad[RW],
\vspace*{2pt}\cr
\displaystyle\delta+(pD+\pi_p)*G_p, &\quad[SAW],
\vspace*{2pt}\cr
\displaystyle\pi_p+\pi_p*p \bigl(D-D(o)\delta
\bigr)*G_p, & \quad[Ising \& percolation].}\hspace*{-20pt}
\end{equation}
Repeated use of these identities yields\footnote{For SAW, since
$\|\pi_p\|_1=o(1)$ as $d\vee L\to\infty$ and $\|G_p\|_\infty<\infty$
for every $p\le p_{\mathrm{c}}$ \cite{h08,hhs03},
\begin{eqnarray*}
G_p&=&\delta+pD*G_p+\pi_p*
\underbrace{G_p}_{\mathrm{replace}}
\\
&=&\delta+pD*G_p+\pi_p* (\delta+pD*G_p+
\pi_p*G_p )
\\
&=&(\delta+\pi_p)+(\delta+\pi_p)*pD*G_p+
\pi_p^{*2}* \underbrace{G_p}_{\mathrm{replace}}
=\cdots\to\mbox{(\ref{eq-lace-exp})}.
\end{eqnarray*}
For percolation and the Ising model, since $D(o)=o(1)$ and
$p\|\pi_p\|_1=1+o(1)$ as $d\vee L\to\infty$ and $\|G_p\|_\infty\le1$
for every $p\le p_{\mathrm{c}}$ \cite{h08,hhs03,s07},
\begin{eqnarray*}
G_p&=&\pi_p+\pi_p*pD*G_p-pD(o)
\pi_p* \underbrace{G_p}_{\mathrm{replace}}
\\
&=&\pi_p+\pi_p*pD*G_p-pD(o)
\pi_p* \bigl(\pi_p+\pi_p*pD*G_p-pD(o)
\pi_p*G_p \bigr)
\\
&=& \bigl(\pi_p-pD(o)\pi_p^{*2} \bigr)+
\bigl(\pi_p-pD(o)\pi_p^{*2}
\bigr)*pD*G_p + \bigl(-pD(o) \bigr)^2
\pi_p^{*2}* \underbrace{G_p}_{\mathrm{replace}}
\\
&=&\cdots\to\mbox{(\ref{eq-lace-exp})}.
\end{eqnarray*}\vspace*{-20pt}}
%
%e1.12 #&#
\begin{equation}
\label{eq-lace-exp} G_p=1_p+\varPi_p*pD*G_p,
\end{equation}
where
%
%e1.13 #&#
\begin{eqnarray}
\label{eq-Pi-def} \varPi_p(x)= \cases{ \delta_{o,x},\qquad\mbox{[RW]},
\vspace*{3pt}\cr
\displaystyle\sum_{n=0}^\infty
\pi_p^{*n}(x)\equiv\sum_{n=0}^\infty
(\underbrace{\pi_p*\cdots*\pi_p}_{n\mbox{-}\mathrm{fold}})(x),
\cr
\hspace*{44pt}\mbox{[SAW],}
\hspace*{5pt}\cr
\displaystyle\sum_{n=1}^\infty
\bigl(-pD(o) \bigr)^{n-1}\pi_p^{*n}(x),
\cr
\hspace*{44pt}\mbox{[Ising \& percolation]}}
\end{eqnarray}
with the convention $f^{*0}(x)\equiv\delta_{o,x}$ for general $f$.
When $d>d_{\mathrm{c}}$ and $d\vee L\gg1$, there is a $\rho>0$ such that
$|\varPi_{p_{\mathrm{c}}}(x)|$ is summable and decays as
$|x|^{-d-2-\rho}$
\mbox{\cite{h08,hhs03,s07}.}
The multiplicative constant $A$ in (\ref{eq-Gpc-fr}) and $p_{\mathrm
{c}}$ can be
represented in terms of $\varPi_{p_{\mathrm{c}}}(x)$ as
%
%e1.14 #&#
\begin{equation}
\label{eq-pcA-fr} \qquad p_{\mathrm{c}}= \biggl(\sum_{x\in\mathbb{Z}^d}
\varPi_{p_{\mathrm
{c}}}(x) \biggr)^{-1},\qquad A=p_{\mathrm{c}}
\biggl(1+ \frac{p_{\mathrm{c}}}{\sigma^2}\sum_{x\in
\mathbb{Z}^d}|x|^2
\varPi_{p_{\mathrm{c}}
}(x) \biggr).
\end{equation}

In this paper, we investigate long-range SAW, percolation and the Ising model
on $\mathbb{Z}^d$ defined by power-law decaying pair potentials of the form
$D(x)\asymp|x|^{-d-\alpha}$ with $\alpha>0$. For example, as in
\cite{csI,csII},
we can consider the following uniformly spread-out long-range $D$ with
parameter $L\in[1,\infty)$:
%
%e1.15 #&#
\begin{equation}
\label{eq-D-exact} D(x)=\frac{\veee{{x}/L}_1^{-d-\alpha}}{\sum_{y\in
\mathbb
{Z}^d}\veee{{y}/L}_1^{-d
-\alpha}},
\end{equation}
where $\veee{x}_\ell=|x|\vee\ell$. As a result,
%
%e1.16 #&#
\begin{equation}
\label{eq-D-def} D(x)=O \bigl(L^\alpha \bigr)\veee{x}_L^{-d-\alpha},
\end{equation}
which we require throughout the paper (cf., Assumption~\ref{assumptiond}
below). The goal is to see how the asymptotic expression (\ref{eq-Gpc-fr})
of $G_{p_{\mathrm{c}}}(x)$ changes depending on the value of $\alpha
$. We note that
(\ref{eq-Gpc-fr}) and (\ref{eq-pcA-fr}) are invalid for $\alpha\le
2$ because then
$\sigma^2=\infty$.

Let
%
%e1.17 #&#
\begin{eqnarray}
\label{eq-dc} d_{\mathrm{c}}= \cases{ 2(\alpha\wedge2), &\quad[SAW \&
Ising],
\vspace*{2pt}\cr
3(\alpha\wedge2), &\quad[percolation].}
\end{eqnarray}
It has been proved \cite{hhs08} that, for $d>d_{\mathrm{c}}$ and
$L\gg1$, the Fourier
transform $\hat G_p(k)\equiv\sum_{x\in\mathbb{Z}^d}e^{ik\cdot
x}G_p(x)$ for the
long-range models is bounded above and below by a~multiple of
$\hat G_{\hat p}^\rw(k)\equiv(1-\hat p\hat D(k))^{-1}$ with $\hat
p=p/p_{\mathrm{c}}$,
uniformly in~$p<p_{\mathrm{c}}$. Although this gives an impression of
the similarity
between $G_{p_{\mathrm{c}}}(x)$ and $G_1^\rw(x)$, it is still too
weak to
identify the
asymptotic expression of $G_{p_{\mathrm{c}}}(x)$. The proof of the
above Fourier-space result makes use of the following properties of $D$
that we make use of here
as well: there are $v_\alpha=O(L^{\alpha\wedge2})$ and $\varepsilon>0$
such that
%
%e1.18 #&#
\begin{eqnarray}
\label{eq-hatDasy} \hat D(k)&\equiv&\sum_{x\in\mathbb{Z}^d}e^{ik\cdot
x}D(x)
\nonumber\\[-8pt]\\[-8pt]
&=& 1-v_\alpha |k|^{\alpha
\wedge2} \times\cases{ \displaystyle1+O \bigl((L|k|)^\varepsilon
\bigr), &\quad$[\alpha\ne2]$,
\vspace*{4pt}\cr
\displaystyle\log\frac{1}{L|k|}+O(1), &
\quad$[ \alpha=2]$.}\nonumber
\end{eqnarray}
If $\alpha>2$, then $v_\alpha=\sigma^2/(2d)$. Moreover, if $L\gg1$, then
there is a constant $\Delta\in(0,1)$ such that\footnote{In the
proof of the bound on $\|D^{*n}\|_\infty$, we simply bounded the factor
$\log\frac\pi{2r}$ in \cite{csI}, (A.4), by a~positive constant. If
we make
the most of that factor instead, we can readily improve the bound for
$\alpha=2$ as
%
%e1.19 #&#
\begin{equation}
\label{eq-Dbdalpha=2} \bigl\|D^{*n}\bigr\|_\infty\le O \bigl(L^{-d}
\bigr) (n\log n)^{-d/2}.
\end{equation}
}
%
%e1.20 #&#
\begin{eqnarray}
\label{eq-Dbd} &\bigl\|D^{*n}\bigr\|_\infty\le O
\bigl(L^{-d} \bigr) n^{-d/(\alpha\wedge2)}\qquad[n\ge1],&
\nonumber\\[-8pt]\\[-8pt]
&1-\hat D(k) \cases{ <2-\Delta, &\quad$ \bigl[k\in[-\pi,
\pi]^d \bigr]$,
\vspace*{4pt}\cr
>\Delta, &\quad$ \bigl[\|k\|_\infty\ge
L^{-1} \bigr]$.}&\nonumber
\end{eqnarray}
All those properties hold for $D$ in (\ref{eq-D-exact}) (cf., \cite{csI,csII,csIII}).

%s1.2 #&#
\subsection{Main result}
In addition to the above properties, the $n$-step transition probability
obeys the following bound:
%
%e1.21 #&#
\begin{eqnarray}
\label{eq-HK1} D^{*n}(x)\le\frac{O(L^{\alpha\wedge2})}{\veee
{x}_L^{d+\alpha
\wedge2}} n\times\cases{ 1, &
\quad$[\alpha\ne2]$,
\vspace*{2pt}\cr
\log\veee{x}_L, &\quad$[\alpha=2]$.}
\end{eqnarray}
This is due to the following two facts: (i)~the contribution
from the walks that have at least one step which is longer than $c\veee{x}_L$
for a given $c>0$ is bounded by $O(L^\alpha)n/\veee{x}_L^{d+\alpha
}$; (ii)~the
contribution from the walks whose $n$ steps are all shorter than
$c\veee{x}_L$
is bounded,\vspace*{1pt} due to the local CLT, by
$O(\tilde vn)^{-d/2}e^{-|x|^2/O(\tilde vn)}\le O(\tilde vn)/\veee{x}_L^{d+2}$
(times an exponentially small normalization constant), where $\tilde v$
is the
variance of the truncated 1-step distribution
$\tilde D(y)\equiv D(y){\mathbh1}_{\{|y|\le c|x|\}}$ and equals
%
%e1.22 #&#
\begin{eqnarray}
\tilde v=\sum_{y\in\mathbb{Z}^d}|y|^2\tilde D(y)\le
O \bigl(L^{\alpha
\wedge2} \bigr) \times\cases{ \veee{x}_L^{2-\alpha},&
\quad$[\alpha<2]$,
\vspace*{2pt}\cr
\log\veee{x}_L, &\quad$[\alpha=2]$,
\vspace*{2pt}\cr
1, &
\quad$[\alpha>2]$.}
\end{eqnarray}
For $\alpha\ne2$, inequality (\ref{eq-HK1}) is a discrete space--time
version of the heat-kernel bound on the transition density $p_s(x)$ of
an $\alpha$-stable/Gaussian process:
%
%e1.23 #&#
\begin{equation}
\label{eq-HK0} p_s(x)\equiv\int_{\mathbb{R}^d}
\frac{\mathrm{d}^dk}{(2\pi)^d} e^{-ik\cdot
x-s|k|^{\alpha
\wedge2}}\le\frac{O(s)}{|x|^{d+\alpha\wedge2}}.
\end{equation}

In Section~\ref{ssRW-Green}, we will show that the properties (\ref{eq-D-def}),
(\ref{eq-hatDasy}) and (\ref{eq-HK1}) are sufficient to obtain an asymptotic
expression of $G_1^\rw(x)$. However, these properties are not good enough
to fully control error terms arising from convolutions of $D^{*n}(x)$ and
$\varPi_p(x)$ in (\ref{eq-Pi-def}). To overcome this difficulty, we assume
the following bound on the discrete derivative of the $n$-step transition
probability:
%
%e1.24 #&#
%e1.25 #&#
\begin{eqnarray}
\label{eq-HK2}
\biggl|D^{*n}(x)-\frac{D^{*n}(x+y)+D^{*n}(x-y)}2 \biggr|\le\frac
{O(L^{\alpha
\wedge2})\veee{y}_L^2}{\veee{x}_L^{d+\alpha\wedge2+2}} n
\nonumber\\[-9pt]\\[-9pt]
\eqntext{\biggl[|y|\le\dfrac13|x|\biggr].}
\end{eqnarray}

Here is the summary of the properties of $D$ that we use throughout the paper.

%
%as1.1 #&#
\begin{assumption}\label{assumptiond}
The $\mathbb{Z}^d$-symmetric 1-step distribution $D$ satisfies the properties
(\ref{eq-D-def}), (\ref{eq-hatDasy}), (\ref{eq-Dbd}), (\ref{eq-HK1}) and (\ref{eq-HK2}).
\end{assumption}

In \hyperref[aeg123]{Appendix}, we will show that the following $D$ satisfies
all properties in the above assumption:
%
%e1.26 #&#
\begin{equation}
\label{eqeg123} D(x)=\sum_{t\in\mathbb{N}}U_L^{*t}(x)
T_\alpha(t),
\end{equation}
where $U_L$ is in a class of $\mathbb{Z}^d$-symmetric distributions
on $\mathbb{Z}^d\cap[-L,L]^d$, and $T_\alpha$ is
the stable distribution on $\mathbb{N}$ with parameter $\alpha/2\ne1$.

Under the above assumption on $D$, we can prove the following theorem.

%
%th1.2 #&#
\begin{theorem}\label{theoremmain}
Let $\alpha>0$, $\alpha\ne2$ and
%
%e1.27 #&#
\begin{equation}
\gamma_\alpha=\frac{\Gamma((d-\alpha\wedge2)/2)}{2^{\alpha
\wedge
2}\pi^{d/2}
\Gamma((\alpha\wedge2)/2)}
\end{equation}
and assume all properties of $D$ in Assumption~\ref{assumptiond}.
Then, for
RW with $d>\alpha\wedge2$ and any $L\ge1$, and for SAW, percolation
and the
Ising model with $d>d_{\mathrm{c}}$ and $L\gg1$, there are $\mu\in
(0,\alpha\wedge
2)$ and
$A=A(\alpha,d,L)\in(0,\infty)$ ($A\equiv1$ for random walk) such
that, as
$|x|\to\infty$,
%
%e1.28 #&#
\begin{equation}
\label{eq-main} G_{p_{\mathrm{c}}}(x)=\frac{\gamma_\alpha/v_\alpha
}{A|x|^{d-\alpha
\wedge
2}}+\frac{O(L^{-
\alpha\wedge2+\mu})}{|x|^{d-\alpha\wedge2+\mu}}.
\end{equation}
As a result, by \cite{hhs08}, $\chi_p$ and $\theta_p$ exhibit the mean-field
behavior (\ref{eq-MF}). Moreover, $p_{\mathrm{c}}$~and~$A$ can
be expressed in term of $\varPi_p$ in (\ref{eq-Pi-def}) as
%
%e1.29 #&#
\begin{eqnarray}
\label{eq-pcA} \qquad p_{\mathrm{c}}=\hat\varPi_{p_{\mathrm{c}}}(0)^{-1},
\qquad A=p_{\mathrm{c}}+ \cases{ 0, &\quad$[\alpha<2]$,
\vspace*{5pt}\cr
\displaystyle \frac{p_{\mathrm{c}}^2}{\sigma^2}\sum_x|x|^2\varPi
_{p_{\mathrm{c}}}(x), &\quad$[\alpha>2]$.}
\end{eqnarray}
\end{theorem}

%
%re1.3 #&#
\begin{remark}
(a)~The finite-range models are formally considered as the \mbox{$\alpha=\infty$}
model. Indeed, the leading term in (\ref{eq-main}) for $\alpha>2$ is
identical to (\ref{eq-Gpc-fr}).

(b)~Following the argument in \cite{h08,s07}, we can ``almost'' prove
Theorem~\ref{theoremmain} for $\alpha>2$ without assuming the bounds on
$D^{*n}(x)$. The shortcoming is the restriction $d>10$, not $d>6$, for
percolation. This is due to the peculiar diagrammatic estimate in \cite{h08},
which we do not use in this paper.

(c)
The asymptotic behavior of $G_{p_{\mathrm{c}}}(x)$ in (\ref{eq-Gpc-fr}) or (\ref{eq-main}) is
a key element for the so-called 1-arm exponent to take on its mean-field
value \cite{hhs??,hhh11,kn11,s04}. For finite-range critical percolation,
for example, the probability that $o\in\mathbb{Z}^d$ is connected to
the surface of
the $d$-dimensional ball of radius $r$ centered at $o$ is bounded above and
below by a multiple of $r^{-2}$ in high dimensions \cite{kn11}. The value
of the exponent may change in a peculiar way depending on the value of
$\alpha$ \cite{hhh11}.

(d)
As described in (\ref{eq-pcA}), the constant $A$ exhibits crossover between
$\alpha<2$ and $\alpha>2$; in particular, $A=p_{\mathrm{c}}$ for
$\alpha<2$ [cf.,
(\ref{eq-bnabla-def}) below]. According to some rough computations, it
seems that
the asymptotic expression of $G_{p_{\mathrm{c}}}(x)$ for $\alpha=2$
is a~mixture
of those
for $\alpha<2$ and $\alpha>2$, with a logarithmic correction:
%
%e1.30 #&#
\begin{equation}
G_{p_{\mathrm{c}}}(x)\underset{|x|\to\infty}\sim\frac{\gamma
_2/v_2}{p_{\mathrm{c}}
|x|^{d-2}\log|x|}.
\end{equation}
One of the obstacles to prove this conjecture is a lack of good control on
convolutions of the RW Green's function and the lace-expansion coefficients
for $\alpha=2$. As hinted in the above expression, we may have to deal with
logarithmic factors more actively than ever. We are currently working
in this
direction.
\end{remark}

%s1.3 #&#
\subsection{Notation and the organization}
From now on, we distinguish $G_p^\rw$ from~$G_p$ for the other three models,
and define
%
%e1.31 #&#
\begin{equation}
S_p=G_p^\rw.
\end{equation}
Here, and in the remainder of the paper, the spatial variables are sometimes
omitted. For example,
%
%e1.32 #&#
\begin{equation}
\label{eq-RW-lace} S_p=\delta+pD*S_p
\end{equation}
is the abbreviated version of the convolution equation
%
%e1.33 #&#
\begin{equation}
S_p(x)=\delta_{o,x}+(pD*S_p) (x)=
\delta_{o,x}+\sum_{y\in\mathbb
{Z}^d}pD(y)
S_p (x-y).
\end{equation}
We also recall the notation
%
%e1.34 #&#
\begin{equation}
\veee{x}_\ell=|x|\vee\ell.
\end{equation}

The remainder of the paper is organized as follows.
In Section~\ref{spreliminaries}, we prove the asymptotic expression
(\ref{eq-main}) for $S_1$, as well as bounds on $S_p$ for $p\le1$ and some
basic properties of $G_p$ for $p\le p_{\mathrm{c}}$. Then, by using
these facts and
the diagrammatic bounds on the lace-expansion coefficients in
\cite{hhs03,s07}, we prove (\ref{eq-main}) for $G_{p_{\mathrm{c}}}$ in
Section~\ref{sproof}.

%s2 #&#
\section{Preliminaries}\label{spreliminaries}
In this section, we derive the asymptotic expression (\ref{eq-main})
for $S_1$,
which will be restated as Proposition~\ref{propositionS}, and prove some
properties of $G_p$ that will be used to prove Theorem~\ref{theoremmain}
in Section~\ref{sproof}.

%s2.1 #&#
\subsection{Asymptotics of $S_p$}\label{ssRW-Green}
%
%pr2.1 #&#
\begin{proposition}\label{propositionS}
Let $\alpha>0$, $\alpha\ne2$ and $d>\alpha\wedge2$, and assume all
properties
but (\ref{eq-HK2}) in Assumption~\ref{assumptiond}. Then there is
a $\mu\in(0,\alpha\wedge2)$ such that, for any $L\ge1$, $p\le1$ and
$\kappa>0$,
%
%e2.1 #&#
%e2.2 #&#
\begin{eqnarray}
\delta_{o,x}&\le& S_p(x)\le\delta_{o,x}+
\frac{O(L^{-\alpha\wedge2})} {
\veee{x}_L^{d-\alpha\wedge2}}\qquad \bigl[\forall x\in\mathbb{Z}^d
\bigr],\label{eq-Spubd}
\\
S_1(x)&=&\frac{\gamma_\alpha/v_\alpha}{|x|^{d-\alpha\wedge
2}}+\frac
{O(L^{-\alpha
\wedge2+\mu})}{|x|^{d-\alpha\wedge2+\mu}}\qquad
\bigl[|x|>L^{1+\kappa
} \bigr],\label{eq-S1asy}
\end{eqnarray}
where a constant in the $O(L^{-\alpha\wedge2+\mu})$ term depends on
$\kappa$.
\end{proposition}

\begin{pf}
Inequality (\ref{eq-Spubd}) is an immediate result of (\ref{eq-RW-lace}),
$p\le1$ and (\ref{eq-Dbd})--(\ref{eq-HK1}) as\footnote{For $\alpha
=2$, we can
readily bound $S_p(x)-\delta_{o,x}$ by using (\ref{eq-Dbdalpha=2}) for
$n\ge N_x\equiv\veee{x}_L^2/(L^2\log\veee{x}_L)$ and (\ref{eq-HK1}) for
$n<N_x$ as
\[
S_p(x)-\delta_{o,x}\le\sum_{n=1}^{N_x-1}D^{*n}(x)+
\sum_{n=N_x}^\infty D^{*n}(x) \le
\frac{O(L^{-2})}{\veee{x}_L^{d-2}\log\veee{x}_L}.
\]
}
%
%e2.3 #&#
\begin{eqnarray}
0&\le& S_p(x)-\delta_{o,x}\nonumber
\\
&\le&\sum_{n=1}^\infty D^{*n}(x)
\nonumber\\[-8pt]\\[-8pt]
&\le& \frac{O(L^{\alpha\wedge2})}{\veee{x}_L^{d+\alpha\wedge2}} \sum_{n=1}^{(\veee{x}_L/L)^{\alpha\wedge2}}n
+O \bigl(L^{-d} \bigr)\sum_{n=(\veee{x}_L/L)^{\alpha\wedge2}}^\infty
n^{-d/(\alpha\wedge2)}\nonumber
\\
&=&\frac{O(L^{-\alpha\wedge2})}{\veee{x}_L^{d-\alpha\wedge
2}}.\nonumber
\end{eqnarray}

To prove the asymptotic expression (\ref{eq-S1asy}), we first rewrite $S_1(x)$
for $d>\alpha\wedge2$~as
%
%e2.4 #&#
\begin{eqnarray}
S_1(x)&=&\int_{[-\pi,\pi]^d}\frac{\mathrm{d}^dk}{(2\pi)^d}
\frac
{e^{-ik\cdot x}} {
1-\hat D(k)}\nonumber
\\
&=&\int_0^\infty\mathrm{d}t\int
_{[-\pi,\pi]^d}\frac{\mathrm{d}^dk} {
(2\pi)^d} e^{-ik\cdot x-t(1-\hat D(k))}
\\
&=&\int_T^\infty\mathrm{d}t\int
_{[-\pi,\pi]^d}\frac{\mathrm
{d}^dk}{(2\pi
)^d} e^{-ik\cdot x-t(1-\hat D(k))}+I_1
\nonumber
\end{eqnarray}
for any $T\in(0,\infty)$, where
%
%e2.5 #&#
\begin{eqnarray}
I_1&=&\int_0^T\mathrm{d}t\int
_{[-\pi,\pi]^d}\frac{\mathrm
{d}^dk}{(2\pi
)^d} e^{-ik
\cdot x-t(1-\hat D(k))}
\nonumber\\[-8pt]\\[-8pt]
&=&\int_0^T\mathrm{d}t\,e^{-t}\sum
_{n=0}^\infty\frac
{t^n}{n!}D^{*n}(x).
\nonumber
\end{eqnarray}
Next, we rewrite the large-$t$ integral as
%
%e2.6 #&#
\begin{equation}
\qquad \int_T^\infty\mathrm{d}t\int_{[-\pi,\pi]^d}
\frac{\mathrm
{d}^dk}{(2\pi)^d} e^{-ik\cdot x-t(1-\hat D(k))}=\int_0^\infty\mathrm{d}t\,p_{v_\alpha t}(x) +\sum_{j=2}^5I_j,
\end{equation}
where $p_s(x)$ is the transition density of an $\alpha$-stable/Gaussian
process [cf., (\ref{eq-HK0})], and for any $R\in(0,\pi)$,
%
%e2.7 #&#
%e2.8 #&#
%e2.9 #&#
%e2.10 #&#
\begin{eqnarray}
I_2&=&-\int_0^T\mathrm{d}t\,p_{v_\alpha t}(x)\equiv-\int_0^T\mathrm
{d}t\int_{\mathbb{R}^d} \frac{\mathrm{d}^dk}{(2\pi)^d} e^{-ik\cdot
x-v_\alpha t|k|^{\alpha
\wedge
2}},
\\
I_3&=&\int_T^\infty\mathrm{d}t\int
_{|k|\le R}\frac{\mathrm
{d}^dk}{(2\pi)^d} e^{-ik\cdot x}
\bigl(e^{-t(1-\hat D(k))}-e^{-v_\alpha t|k|^{\alpha
\wedge2}} \bigr),
\\
I_4&=&\int_T^\infty\mathrm{d}t\int
_{[-\pi,\pi]^d}\frac{\mathrm
{d}^dk}{(2\pi
)^d} e^{-ik\cdot x-t(1-\hat D(k))}{
\mathbh1}_{\{|k|>R\}},
\\
I_5&=&-\int_T^\infty\mathrm{d}t\int
_{|k|>R}\frac{\mathrm
{d}^dk}{(2\pi)^d} e^{-ik\cdot x-v_\alpha t|k|^{\alpha\wedge2}}.
\end{eqnarray}
By using the identity
%
%e2.11 #&#
\begin{eqnarray}
&& \int_0^\infty\mathrm{d}t\,e^{-v_\alpha t|k|^{\alpha\wedge2}}
\nonumber\\[-8pt]\\[-8pt]
&&\qquad = \frac{1}{v_\alpha|k|^{\alpha\wedge2}}
=\frac{1}{v_\alpha\Gamma((\alpha\wedge2)/2)} \int_0^\infty\mathrm{d}t\,t^{((\alpha\wedge2)/2)-1}e^{-|k|^2t},\nonumber
\end{eqnarray}
we obtain
%
%e2.12 #&#
\begin{eqnarray}
&& \int_0^\infty\mathrm{d}t\,p_{v_\alpha t}(x)\nonumber
\\
&&\qquad = \frac{1}{v_\alpha\Gamma((\alpha\wedge2)/2)}\int_0^\infty\mathrm{d}t\,t^{((\alpha\wedge2)/2)-1}\int_{\mathbb{R}^d} \frac{\mathrm{d}^dk}{(2\pi)^d} e^{-|k|^2t-ik\cdot x}
\\
&&\qquad =\frac{\gamma_\alpha/v_\alpha}{|x|^{d-\alpha\wedge2}}.\nonumber
\end{eqnarray}
As a result, we arrive at
%
%e2.13 #&#
\begin{equation}
\label{eq-main+error} S_1(x)=\frac{\gamma_\alpha/v_\alpha}{|x|^{d-\alpha
\wedge2}}+\sum
_{j=1}^5I_j.
\end{equation}

It remains to estimate $\sum_{j=1}^5I_j$. First, by (\ref{eq-HK1}) and
(\ref{eq-HK0}), we can estimate $I_1+I_2$ for $|x|>L$ as
%
%e2.14 #&#
\begin{equation}
\label{eq-I1I2-prebd} |I_1+I_2|\le\frac{O(L^{\alpha\wedge
2})}{|x|^{d+\alpha\wedge
2}}\int
_0^T \mathrm{d}t\,t\le\frac{O(L^{\alpha\wedge2})T^2}{|x|^{d+\alpha
\wedge2}}.
\end{equation}
Let
%
%e2.15 #&#
\begin{equation}
\label{eq-muT} \mu=\frac{2(\alpha\wedge2)\varepsilon}{d+\alpha\wedge
2+\varepsilon},\qquad T= \biggl(\frac{|x|}L
\biggr)^{\alpha\wedge2-\mu/2}.
\end{equation}
Then we obtain
%
%e2.16 #&#
\begin{equation}
\label{eq-I1I2-bd} |I_1+I_2|\le\frac{O(L^{-\alpha\wedge2+\mu
})}{|x|^{d-\alpha\wedge
2+\mu}}.
\end{equation}

Next, we estimate $I_3$. For small $R$, whose value will be determined
shortly, we use (\ref{eq-hatDasy}) to obtain
%
%e2.17 #&#
\begin{equation}
\bigl|e^{-t(1-\hat D(k))}-e^{-v_\alpha t\bigr|k|^{\alpha\wedge2}} |\le O \bigl(L^{\alpha\wedge2+\varepsilon}
\bigr)t|k|^{\alpha\wedge2+\varepsilon
}e^{-v_\alpha t
|k|^{\alpha\wedge2}}.
\end{equation}
Therefore, by (\ref{eq-muT}),
%
%e2.18 #&#
\begin{eqnarray}
\label{eq-I3-bd} |I_3|&\le& O \bigl(L^{\alpha\wedge2+\varepsilon} \bigr)\int
_T^\infty\mathrm{d}t\,t\int_{|k|\le R}\,\mathrm{d}^dk |k|^{\alpha\wedge2+\varepsilon}e^{-v_\alpha
t|k|^{\alpha\wedge 2}}\nonumber
\\
&=&O \bigl(L^{\alpha\wedge2+\varepsilon} \bigr)\int_T^\infty\mathrm{d}t\,t\int_0^{v_\alpha t
R^{\alpha\wedge2}}\frac{\mathrm{d}r}r
\biggl(\frac{r}{v_\alpha t} \biggr)^{(d +\alpha\wedge2+\varepsilon)/(\alpha\wedge2)}e^{-r}
\nonumber\\[-8pt]\\[-8pt]
&\le& O \bigl(L^{\alpha\wedge2+\varepsilon} \bigr)\int_T^\infty\mathrm{d}t\,t (v_\alpha t)^{-(d+\alpha\wedge2+\varepsilon)/(\alpha \wedge 2)} \nonumber
\\
&\le& O \bigl(L^{-d} \bigr)T^{1-((d+\varepsilon)/(\alpha
\wedge2))}=\frac{O(L^{-\alpha\wedge2+\mu})}{|x|^{d-\alpha\wedge 2+\mu}}.\nonumber
\end{eqnarray}

Finally we estimate $I_4+I_5$ and determine the value of $R$ during the
course. First, by (\ref{eq-hatDasy})--(\ref{eq-Dbd}), we have
%
%e2.19 #&#
\begin{eqnarray}
\label{eq-I4-prebd} \qquad |I_4|&\le&\int_T^\infty\mathrm{d}t\int_{[-\pi,\pi]^d}\frac
{\mathrm
{d}^dk}{(2\pi)^d} e^{-t(1-\hat D(k))}{
\mathbh1}_{\{|k|>R\}}\nonumber
\\
&&\hspace*{66.5pt}{} \times({\mathbh1}_{\{\|k\|_\infty<L^{-1}\}}+{\mathbh1}_{\{
\|k\|_\infty\ge L^{-1}\}} )
\nonumber\\[-8pt]\\[-8pt]
&\le& \int_T^\infty\mathrm{d}t \biggl(\int
_{|k|>R}\frac{\mathrm{d}^dk}{(2\pi
)^d} e^{-tc(L|k|)^{\alpha\wedge2}}+O(1)e^{-t\Delta}
\biggr)
\nonumber
\\
&\le& O \bigl(L^{-d} \bigr)\int_T^\infty\mathrm{d}t\,t^{-d/(\alpha\wedge
2)}\Gamma \biggl(\frac{d}{\alpha\wedge2};tc(LR)^{\alpha\wedge2}
\biggr)+O(1)e^{-T\Delta},
\nonumber
\end{eqnarray}
where\vspace*{1pt} $\Gamma(a;x)\equiv\int_x^\infty\mathrm{d}t\,t^{a-1}e^{-t}$ is the
incomplete gamma function, which is bounded by $O(x^{a-1})e^{-x}$ for
large $x$.
Here, we choose $R$ to satisfy
%
%e2.20 #&#
\begin{equation}
tc(LR)^{\alpha\wedge2}=\frac{2\varepsilon}{\alpha\wedge2}\log t.
\end{equation}
Then, for large $t$,
%
%e2.21 #&#
\begin{eqnarray}
&& \Gamma \biggl(\frac{d}{\alpha\wedge2};tc(LR)^{\alpha\wedge2} \biggr)\nonumber
\\
&&\qquad \le O \bigl(\bigl(tc(L R)^{\alpha\wedge2} \bigr)^{(d/(\alpha\wedge2))-1} \bigr)e^{-tc(LR)^{\alpha
\wedge2}}
\\
&&\qquad = O \bigl((\log t)^{(d/(\alpha\wedge2))-1} \bigr)t^{-(2\varepsilon)/(\alpha\wedge2)}\le O
\bigl(t^{-\varepsilon/(\alpha\wedge2)} \bigr).\nonumber
\end{eqnarray}
Therefore, again by (\ref{eq-muT}) [cf., (\ref{eq-I3-bd})],
%
%e2.22 #&#
\begin{eqnarray}
\label{eq-I4-1stbd}
&& O \bigl(L^{-d} \bigr)\int_T^\infty\mathrm{d}t\,t^{-d/(\alpha\wedge 2)}\Gamma \biggl(\frac{d} {
\alpha\wedge2};tc(LR)^{\alpha\wedge2}
\biggr)
\nonumber\\[-8pt]\\[-8pt]
&&\qquad \le O \bigl(L^{-d} \bigr)T^{1-((d+\varepsilon)/(\alpha\wedge2))}
=\frac{O(L^{-\alpha\wedge2+\mu})}{|x|^{d-\alpha\wedge2+\mu}}.\nonumber
\end{eqnarray}

We can estimate $I_5$ in exactly the same way.
The exponentially decaying term in (\ref{eq-I4-prebd}) obeys the same bound,
since, for sufficiently large $N$ (depending on $\kappa$),
%
%e2.23 #&#
\begin{eqnarray}
e^{-T\Delta}&\le&\frac{{}^\exists c_N}{T^N}=c_NL^{-d}T^{1-((d+\varepsilon)/(\alpha\wedge2))}L^dT^{-(N+1-((d+\varepsilon)/(\alpha\wedge2)))}\nonumber
\\
&\le& c_NL^{-d}T^{1-((d+\varepsilon)/(\alpha\wedge2))}L^{d-(N+1-((d+\varepsilon)/(\alpha\wedge2)))(\alpha\wedge2-\mu/2)\kappa}
\nonumber\\[-8pt]\\[-8pt]
&&\bigl[\because |x|>L^{1+\kappa}\Rightarrow T>L^{(\alpha\wedge2-\mu/2)\kappa} \bigr]\nonumber
\\
&\le& c_NL^{-d}T^{1-((d+\varepsilon)/(\alpha\wedge2))}=\frac
{O(L^{-\alpha\wedge2+\mu})}{|x|^{d-\alpha\wedge2+\mu}}.\nonumber
\end{eqnarray}

Summarizing the above, we obtain that, for $|x|>L^{1+\kappa}$,
%
%e2.24 #&#
\begin{equation}
\label{eq-error-est} \Biggl|\sum_{j=1}^5I_j
\Biggr|\le\frac{O(L^{-\alpha\wedge2+\mu
})}{|x|^{d-\alpha
\wedge2+\mu}}.
\end{equation}
This together with (\ref{eq-main+error}) completes the proof of
Proposition~\ref{propositionS}.
\end{pf}

%s2.2 #&#
\subsection{Basic properties of $G_p$}\label{ssbasic-props}
In this subsection, we summarize some basic properties of $G_p$.
Roughly speaking, those properties are the continuity up to
\mbox{$p=p_{\mathrm{c}}$}
(Lemma~\ref{lemmaGcont}), the RW bound that is optimal for $p\le1$
(Lemma~\ref{lemmaRWbds}) and the \textit{a priori} bound that is not sharp
but finite as long as $p<p_{\mathrm{c}}$ (Lemma~\ref{lemmasubpcbd}).
We will use
them in the next section (especially in Section~\ref{ssopt-bound}) to
prove Theorem~\ref{theoremmain}.

%
%le2.2 #&#
\begin{lemma}\label{lemmaGcont}
For every $x\in\mathbb{Z}^d$, $G_p(x)$ is nondecreasing and
continuous in $p<p_{\mathrm{c}}
$ for
\textup{SAW}, and in $p\le p_{\mathrm{c}}$ for percolation and the Ising model. The
continuity up
to $p=p_{\mathrm{c}}^\saw$ for \textup{SAW} is also valid if $G_p^\saw(x)$ is
uniformly
bounded in \mbox{$p<p_{\mathrm{c}}^\saw$}.
\end{lemma}

\begin{pf}
For SAW, since $G_p^\saw(x)$ is a power series of $p\ge0$ with nonnegative
coefficients, it is nondecreasing and continuous in $p<p_{\mathrm
{c}}^\saw$.
The continuity up to $p=p_{\mathrm{c}}^\saw$ under the hypothesis is
due to monotone
convergence.

For the Ising model, we first note that, by Griffiths' inequality
\cite{g70}, $\langle\varphi_o\varphi_x\rangle_{\beta,\Lambda}$
is nondecreasing and
continuous in $\beta\ge0$ and nondecreasing in $\Lambda\subset
\mathbb{Z}^d$.
Therefore, the infinite-volume limit $G_p^\ising(x)$ is nondecreasing and
left-continuous in $p\ge0$. The continuity in $p\le p_{\mathrm
{c}}^\ising$
follows from
the fact that, for $p<p_{\mathrm{c}}^\ising$, $G_p^\ising(x)$
coincides with the
decreasing limit of the finite-volume two-point function under the
``plus-boundary'' condition, which is right-continuous in $p\ge0$.

For percolation, $G_p^\perc(x)$ is nondecreasing in $p\ge0$ because
the event
that there is a path of occupied bonds from $o$ to $x$ is an increasing event.
The continuity in $p\ge0$ is obtained by following the same strategy as
explained above for the Ising model and using the fact that there is at most
one infinite occupied cluster for all $p\ge0$. This completes the
proof of
Lemma~\ref{lemmaGcont}.
\end{pf}

%
%le2.3 #&#
\begin{lemma}\label{lemmaRWbds}
For every $p<p_{\mathrm{c}}$ and $x\in\mathbb{Z}^d$,
%
%e2.25 #&#
\begin{eqnarray}
\label{eq-RWbds} G_p(x)&\le& S_p(x),\qquad pD(x) (1-
\delta_{o,x})\le G_p(x)-\delta_{o,x}\le
(pD*G_p) (x).\hspace*{-35pt}
\end{eqnarray}
\end{lemma}

\begin{pf}
The first inequality for $p>1\equiv p_{\mathrm{c}}^\rw$ is trivial since
$S_p(x)=\infty$
for every $x\in\mathbb{Z}^d$. On the other hand, the first inequality
for $p\le
1$ is
obtained by using the second inequality $N$ times and then using (\ref{eq-Dbd}),
as
%
%e2.26 #&#
\begin{eqnarray}
G_p(x)&\le&\sum_{n=0}^{N-1}p^nD^{*n}(x)+p^N
\bigl(D^{*N}*G_p \bigr) (x)
\nonumber\\[-8pt]\\[-8pt]
&\le& S_p(x)+\bigl\|D^{*N}\bigr\|_\infty
\chi_p \underset{N\uparrow\infty}\to S_p(x).\nonumber
\end{eqnarray}

It remains to prove the second inequality in (\ref{eq-RWbds}). In
fact, it
suffices to prove the inequality only for $x\ne o$, since $G_p(o)=1$ for
all three models and therefore the inequality is trivial for $x=o$.
For SAW and percolation, the inequality is obtained by specifying the first
step $pD$ and then using subadditivity for SAW or the BK inequality for
percolation \cite{bk85}. For the Ising model,
we use the following random-current representation \cite{a82,ghs70} (see
also \cite{s07}, Section~2.1):
%
%e2.27 #&#
\begin{equation}
\label{eq-RC-rep} \langle\varphi_o\varphi_x
\rangle_{\beta,\Lambda}=\frac
{\sum_{\partial\mathbf{n}
=\{o\}
\triangle\{x\}}w_\Lambda(\mathbf{n})}{\sum_{\partial
\mathbf{n}
=\varnothing}
w_\Lambda(\mathbf{n})},\qquad w_\Lambda(\mathbf{n})=
\prod_{\{u,v\}\subset\Lambda}\frac{(\beta
J_{u,v})^{n_{u,v}}} {
n_{u,v}!},\hspace*{-30pt}
\end{equation}
where $\mathbf{n}\equiv\{n_{u,v}\}$ is a collection of $\mathbb{Z}_+$-valued
undirected bond
variables (i.e., $n_{u,v}=n_{v,u}\in\mathbb{Z}_+\equiv\{0\}\cup
\mathbb{N}$ for each bond
$\{u,v\}\subset\Lambda$), $\partial\mathbf{n}$ is the set of
vertices $y$ such
that $\sum_{z\in\Lambda}n_{y,z}$ is an odd number, and ``$\triangle$''
represents symmetric difference (i.e.,
$\{o\}\triangle\{x\}=\varnothing$ if $x=o$, otherwise
$\{o\}\triangle\{x\}=\{o,x\}$). Using this
representation, we prove below that, for $x\ne o$,
%
%e2.28 #&#
\begin{equation}
\label{eq-RWbds-Ising} pD(x)\le\langle\varphi_o\varphi_x
\rangle_{\beta,\Lambda}\le\sum_{y\in\Lambda}pD(y) \langle
\varphi_y\varphi_x\rangle_{\beta,\Lambda},
\end{equation}
where $pD(x)=\tanh(\beta J_{o,x})$. The second inequality in (\ref{eq-RWbds})
for the Ising model
is the infinite-volume limit of the above inequality.

To prove the lower bound of (\ref{eq-RWbds-Ising}), we first specify
the parity
of $n_{o,x}$ to obtain that, for $x\ne o$ (so that
$\{o\}\triangle\{x\}=\{o,x\}$),
%
%e2.29 #&#
\begin{equation}
\label{eq-lowerbd-pr1} \langle\varphi_o\varphi_x
\rangle_{\beta,\Lambda}= \frac
{\mathop{\sum_{\partial\mathbf{n}=\{o,x\}, (n_{o,x}\odd)}}w_\Lambda(\mathbf{n})+\mathop
{\sum_{\partial
\mathbf{n}=\{o,x\}, (n_{o,x}\even)}} w_\Lambda(\mathbf
{n})}{
\mathop{\sum_{\partial\mathbf{n}=\varnothing, (n_{o,x}\odd)}}w_\Lambda
(\mathbf{n})
+\mathop{\sum_{\partial\mathbf{n}=\varnothing, (n_{o,x}\even
)}}w_\Lambda
(\mathbf{n})}.\hspace*{-25pt}
\end{equation}
Let
%
%e2.30 #&#
\begin{equation}
\tilde Y_y(z,x)\equiv\mathop{\sum_{\partial\mathbf{n}=\{z\}
\triangle\{x\}}}_{
(n_{o,y}\even)}w_\Lambda(
\mathbf{n}),\qquad \tilde Z_y\equiv\mathop{\sum
_{\partial\mathbf{n}=\varnothing}}_{(n_{o,y}
\even)}w_\Lambda(\mathbf{n}).
\end{equation}
Then, by changing
the parity of $n_{o,x}$ (and the constraint on $\partial\mathbf{n}$
accordingly)
and recalling $\tanh(\beta J_{o,x})=pD(x)$, we obtain
%
%e2.31 #&#
%e2.32 #&#
\begin{eqnarray}
\mathop{\sum_{\partial\mathbf{n}=\{o,x\}}}_{(n_{o,x}\odd
)}w_\Lambda
(\mathbf{n}) &=&pD(x)\mathop{\sum_{\partial\mathbf{n}=\varnothing
}}_{(n_{o,x}\even)}
w_\Lambda(\mathbf{n})=pD(x)\tilde Z_x,\label{eq-lowerbd-pr2}
\\
\mathop{\sum_{\partial\mathbf{n}=\varnothing}}_{(n_{o,x}\odd
)}w_\Lambda
(\mathbf{n})&=&pD(x)\mathop{\sum_{\partial\mathbf{n}=\{o,x\}
}}_{(n_{o,x}\even)}
w_\Lambda(\mathbf{n})=pD(x)\tilde Y_x(o,x),\label{eq-lowerbd-pr3}
\end{eqnarray}
hence
%
%e2.33 #&#
\begin{eqnarray}
\langle\varphi_o\varphi_x\rangle_{\beta,\Lambda}&=&
\frac{pD(x)\tilde Z_x+\tilde Y_x(o,x)} {
pD(x)\tilde Y_x(o,x)+\tilde Z_x}
\nonumber\\[-8pt]\\[-8pt]
&=&pD(x)+\frac{(1-p^2D(x)^2)\tilde Y_x(o,x)}{pD(x)\tilde
Y_x(o,x)+\tilde Z_x}
\ge pD(x).\nonumber
\end{eqnarray}

To prove the upper bound in (\ref{eq-RWbds-Ising}), we first note
that, if
$\partial\mathbf{n}=\{o,x\}$, then there must be at least one $y\in
\Lambda$
such that $n_{o,y}$ is an odd number. By similar computation to
(\ref{eq-lowerbd-pr2}), we obtain that, for $x\ne o$,
%
%e2.34 #&#
\begin{eqnarray}
\label{eq-upperbd-pr1} \sum_{\partial\mathbf{n}=\{o,x\}}w_\Lambda(
\mathbf{n})&\le&\sum_{y\in\Lambda} \mathop{\sum
_{\partial\mathbf{n}=\{o,x\}}}_{(n_{o,y}\odd
)}w_\Lambda(\mathbf{n} )
\nonumber\\[-8pt]\\[-8pt]
&=&\sum_{y\in\Lambda}pD(y)\underbrace{\mathop{\sum
_{\partial
\mathbf{n}=\{y\}
\triangle\{x\}}}_{(n_{o,y}\even)}w_\Lambda(\mathbf{n})}_{\tilde Y_y(y,x)}.
\nonumber
\end{eqnarray}
Moreover, $\tilde Y_y (y, x)\le
\sum_{\partial\mathbf{n}=\{y\} \Delta\{x\}}w_{\Lambda} (\mathbf{n})$ for any $y\in \Lambda$. Therefore, for $x\ne o$,
%
%e2.35 #&#
\begin{eqnarray}
\label{eq-upperbd-pr2} \langle\varphi_o\varphi_x
\rangle_{\beta,\Lambda}&\equiv&\frac{\sum_{\partial
\mathbf{n}
=\{o,x\}}w_\Lambda(\mathbf{n})}{\sum_{\partial\mathbf
{n}=\varnothing
}w_\Lambda
(\mathbf{n})}\nonumber
\\
&\le& \sum
_{y\in\Lambda}\frac{pD(y)\tilde
Y_y(y,x)}{\sum_{\partial \mathbf{n}=\varnothing}w_\Lambda (\mathbf{n})}
\\
&\le&\sum_{y\in\Lambda}pD(y)\langle\varphi_y
\varphi_x\rangle_{\beta,\Lambda}.
\nonumber
\end{eqnarray}
%
%where we have used the fact that $\tilde Y_y(y,x)/\tilde Z_y$ is
%equivalent to
%the finite-volume two-point function under the restriction $J_{o,y}=0$ and
%therefore, by using Griffiths' inequality \cite{g70}, it is bounded above
%by $\langle\varphi_y\varphi_x\rangle_{\beta,\Lambda}$.
This
completes the proof of~(\ref{eq-RWbds-Ising}), hence the proof of Lemma~\ref{lemmaRWbds}.
\end{pf}

%
%le2.4 #&#
\begin{lemma}\label{lemmasubpcbd}
Assume the property (\ref{eq-D-def}) in Assumption~\ref{assumptiond}. Then,
for every $\alpha>0$ and $p<p_{\mathrm{c}}$, there is a
$K_p=K_p(\alpha,d,L)<\infty$
such that, for any $x\in\mathbb{Z}^d$,
%
%e2.36 #&#
\begin{equation}
\label{eq-subpcbd} G_p(x)\le K_p\veee{x}_L^{-d-\alpha}.
\end{equation}
\end{lemma}

%
%re2.5 #&#
\begin{remark}
This together with the lower bound in (\ref{eq-RWbds}) implies
that, for
every $p<p_{\mathrm{c}}$, $G_p(x)$ is bounded above and below by a
$p$-dependent multiple of~$\veee{x}_L^{-d-\alpha}$. This shows sharp contrast to the exponential
decay of $G_p(x)$ for the finite-range models.
\end{remark}

\begin{pf*}{Proof of Lemma~\ref{lemmasubpcbd}}
Since $G_p(o)\le\chi_p<\infty$ for $p<p_{\mathrm{c}}$, it suffices
to prove
(\ref{eq-subpcbd}) for $x\ne o$. We follow the idea of the proof of
\cite{an86}, Lemma~5.2, for one-dimensional long-range percolation and extend
it to those three models in general dimensions. The key ingredient is the
following Simon--Lieb type inequality: for~$0<\ell<|x|$,
%
%e2.37 #&#
\begin{equation}
\label{eq-SimonLieb} G_p(x)\le\mathop{\sum_{\{u,v\}\subset\mathbb
{Z}^d}}_{(|u|\le\ell
<|v|)}G_p(u)
pD(v-u) G_p(x-v).
\end{equation}
For SAW and percolation, this is a result of subadditivity or the BK
inequality (cf., e.g., \cite{g99,ms93}). For the Ising model, this is
obtained by using the random-current representation (\ref{eq-RC-rep}) and
a restricted version of the source-switching lemma~\cite{s07}, Lemma~2.3,
as follows. Let $Z_\Lambda=\sum_{\partial\mathbf{n}=\varnothing
}w_\Lambda
(\mathbf{n})$
such that, for $x\ne o$,
%
%e2.38 #&#
\begin{equation}
\langle\varphi_o\varphi_x\rangle_{\beta,\Lambda}=\sum
_{\partial
\mathbf{n}=\{o,x\}} \frac{w_\Lambda(\mathbf{n})}{Z_\Lambda}.
\end{equation}
We note that, if $\partial\mathbf{n}=\{o,x\}$, then there is a path
$\omega=(\omega_0,\omega_1,\ldots,\omega_t)\subset\Lambda$ from
$\omega_0=o$
to $\omega_t=x$ such that $n_{\omega_{s-1},\omega_s}$ is odd for every
$s\in\{1,\ldots,t\}$; moreover, there is a unique $\tau\in\{1,\ldots,t\}
$ such
that $|\omega_{\tau-1}|\le\ell<|\omega_\tau|$ (i.e., $\tau$ is the
first time
when $\omega$ crosses the surface of the ball $B_\ell$ of radius
$\ell$
centered at the origin). This can be restated as follows: if
$\partial\mathbf{n}=\{o,x\}$, then there is a bond $\{u,v\}\subset
\Lambda$ such
that $n_{u,v}$ is odd and that $u$ is connected from $o$ with a
path of bonds $\subset B_\ell$ with odd numbers. Therefore,
%
%e2.39 #&#
\begin{equation}
\label{eq-SL-step1} \langle\varphi_o\varphi_x
\rangle_{\beta,\Lambda}\le\mathop{\sum_{\{u,v\}
\subset\Lambda}}_{(|u|\le\ell<|v|)}
\sum_{\partial\mathbf{n}=\{
o,x\}}\frac
{w_\Lambda
(\mathbf{n})}{Z_\Lambda} {\mathbh1}_{\{n_{u,v}\odd\}}
{\mathbh1}_{\{o\undersett{\mathbf{n}} \longleftrightarrow u\ \mathrm{in}\
B_\ell\}},
\end{equation}
where $\{o\undersett{\mathbf{n}}\longleftrightarrow u$ in $B_\ell\}$
is the event
that $o$ is connected to $u$ with a path of bonds $b\subset B_\ell$ satisfying
$n_b>0$. Multiplying $Z_{B_\ell}/Z_{B_\ell}\equiv1$ to both sides of
(\ref{eq-SL-step1}) and using the identity
$Z_{B_\ell}=\sum_{\partial\mathbf{m}=\varnothing}w_{B_\ell
}(\mathbf{m})$,
we obtain
%
%e2.40 #&#
\begin{equation}
\langle\varphi_o\varphi_x\rangle_{\beta,\Lambda}\le
\mathop{\sum_{\{u,v\}
\subset
\Lambda}}_{(|u|\le\ell<|v|)}\mathop{\sum
_{\partial\mathbf
{m}=\varnothing}}_{
\partial\mathbf{n}=\{o,x\}}\frac{w_{B_\ell}(\mathbf{m})}{Z_{B_\ell
}}
\frac
{w_\Lambda
(\mathbf{n})}{Z_\Lambda} {\mathbh1}_{\{n_{u,v}\odd\}}
{\mathbh1}_{\{o\undersett{\mathbf{m}+\mathbf
{n}}\longleftrightarrow u\ \mathrm{in}\ B_\ell\}},\hspace*{-35pt}
\end{equation}
where we have used the trivial inequality
${\mathbh1}_{\{o\undersett{\mathbf{n}}\longleftrightarrow u\ \mathrm
{in}\ B_\ell\}}\le
{\mathbh1}_{\{o\undersett{\mathbf{m}+\mathbf{n}}\longleftrightarrow
u\ \mathrm{in}\ B_\ell\}}$.
Then, by using the source-switching lemma
\cite{s07}, Lemma~2.3, we obtain
%
%e2.41 #&#
\begin{eqnarray}
\langle\varphi_o\varphi_x\rangle_{\beta,\Lambda}&\le&
\mathop{\sum_{\{u,v\}
\subset\Lambda}}_{(|u|\le\ell<|v|)}\mathop{\sum
_{\partial\mathbf
{m}=\{o\}
\triangle\{u\}}}_{\partial\mathbf{n}=\{u,x\}}\frac{w_{B_\ell
}(\mathbf{m}
)}{Z_{B_\ell}}
\frac{w_\Lambda(\mathbf{n})}{Z_\Lambda} {\mathbh1}_{\{n_{u,v}\odd
\}}
{\mathbh1}_{\{o \undersett{\mathbf{m}+\mathbf
{n}}\longleftrightarrow u\ \mathrm{in}\ B_\ell\}}\hspace*{-30pt}
\nonumber\\[-7pt]\\[-7pt]
&=&\mathop{\sum_{\{u,v\}\subset\Lambda}}_{(|u|\le\ell<|v|)}\langle
\varphi_o \varphi_u\rangle_{\beta,B_\ell}\mathop{\sum
_{\partial
\mathbf{n}
=\{u,x\}}}_{(n_{u,v}\odd)}\frac{w_\Lambda(\mathbf{n})}{Z_\Lambda
},\hspace*{-30pt}
\nonumber
\end{eqnarray}
where we have used the identity
${\mathbh1}_{\{o\undersett{\mathbf{m}+\mathbf{n}}\longleftrightarrow
u\ \mathrm{in}\ B_\ell\}}=1$ given
$\partial\mathbf{m}=\{o\}\triangle\{u\}$ and then used (\ref{eq-RC-rep}).
Finally, by
following the same argument as in (\ref{eq-upperbd-pr1})--(\ref{eq-upperbd-pr2}) and
then taking the infinite-volume limit, we obtain (\ref{eq-SimonLieb})
for the
Ising model.

Now we prove (\ref{eq-subpcbd}) by using (\ref{eq-SimonLieb}) with
$\ell=\frac{1}3|x|$
(the factor $\frac{1}3$ is unimportant as long as it is less than $\frac{1}2$).
Let
%
%e2.42 #&#
\begin{equation}
c_x=\mathop{\sum_{\{u,v\}\subset\mathbb{Z}^d}}_{(|u|\le(1/3)|x|<|v|)}G_p(u)
D(v-u).
\end{equation}
We note that $c_x\to0$ as $|x|\to\infty$, because
%
%e2.43 #&#
\begin{eqnarray}
c_x&=&\mathop{\sum_{\{u,v\}\subset\mathbb{Z}^d}}_{(|u|\le(1/4)|x|, (1/3)|x|<|v|)}
G_p(u) pD(v-u)\nonumber
\\[1pt]
&&{}+\mathop{\sum_{\{u,v\}\subset\mathbb{Z}^d}}_{((1/4)|x|<|u|\le (1/3)|x|<|v|)}G_p(u)pD(v-u)
\\[1pt]
&\le&\chi_p p\underbrace{\sup_{u\dvtx|u|\le (1/4)|x|}\sum
_{v\dvtx|v|>(1/3)|x|} D(v-u)}_{O(|x|^{-\alpha})}+p\underbrace{\sum
_{u\dvtx|u|>(1/4)|x|}G_p (u)}_{\mathrm{Tail\ of\ }\chi_p<\infty}.\hspace*{-30pt}
\nonumber
\end{eqnarray}
Therefore, for any $\varepsilon\in(0,1)$, there is an $\tilde\ell\in
[L,\infty)$
such that $2^{d+\alpha}c_xp\le\varepsilon$ for all $|x|\ge\tilde\ell$.
Then, for
$|x|\ge\tilde\ell$, (\ref{eq-SimonLieb}) implies
%
%e2.44 #&#
\begin{eqnarray}
\label{eq-SimonLiebAppl}
G_p(x)&\le&\mathop{\sum_{\{u,v\}\subset\mathbb{Z}^d}}_{(|u|\le(1/3)|x|<|v|\le(1/2)|x|)}G_p(u) pD(v-u)G_p(x-v)\nonumber
\\[1pt]
&&{} + \mathop{\sum_{\{u,v\}\subset\mathbb{Z}^d}}_{(|u|\le(1/3)|x|, |v|>(1/2)|x|)} G_p(u) pD(v-u) G_p(x-v)
\\[1pt]
&\le& c_xp\sup_{v\dvtx|v|\le(1/2)|x|}G_p(x-v)+\chi_p^2 p\mathop{\sup_{\{
u,v\}\subset\mathbb{Z}^d}}_{(|u|\le(1/3)|x|, |v|>(1/2)|x|)}D(v-u)\hspace*{-23pt}\nonumber
\\[1pt]
&\le& 2^{-d-\alpha}\varepsilon\sup_{v\dvtx|v|>(1/2)|x|}G_p(v)+
\frac{C_p}{\veee{x}_L^{d+\alpha}}\nonumber
\end{eqnarray}
for some $C_p=O(\chi_p^2)$. If $2\tilde\ell\le|x|<4\tilde\ell$,
then we use (\ref{eq-SimonLiebAppl}) twice to obtain
%
%e2.45 #&#
\begin{eqnarray}
G_p(x)&\le& \bigl(2^{-d-\alpha}\varepsilon \bigr)^2
\sup_{v\dvtx|v|>(1/4)|x|}G_p(v)\nonumber
\\
&&{} +2^{-d-\alpha} \varepsilon \frac{C_p}{\veee{x/2}_L^{d+\alpha}}+\frac{C_p}{\veee
{x}_L^{d+\alpha}}
\\
&=& \bigl(2^{-d-\alpha}\varepsilon \bigr)^2\sup
_{v\dvtx|v|>(1/4)|x|}G_p(v)+(1+ \varepsilon)\frac{C_p} {
\veee{x}_L^{d+\alpha}}.\nonumber
\end{eqnarray}
In general, if $2^{n-1}\tilde\ell\le|x|<2^n\tilde\ell$ for some
$n\in\mathbb{N}
$, then
we repeatedly use (\ref{eq-SimonLiebAppl}) to obtain
%
%e2.46 #&#
\begin{eqnarray}
\label{eq-subpcbd-pr} G_p(x)&\le& \bigl(2^{-d-\alpha}\varepsilon
\bigr)^n\sup_{v\dvtx|v|>(1/2^n)|x|}G_p(v)\nonumber
\\
&&{} + \bigl(1+\varepsilon+  \cdots+\varepsilon^{n-1} \bigr)
\frac{C_p}{\veee{x}_L^{d+\alpha}}
\\
&\le&\frac{\tilde\ell^{d+\alpha}}{\veee{x}_L^{d+\alpha}}\chi_p+\frac
{C_p} {
(1-\varepsilon)\veee{x}_L^{d+\alpha}}.
\nonumber
\end{eqnarray}
For $|x|<\tilde\ell$, we use the trivial inequality
$G_p(x)\le\chi_p\le\tilde\ell^{d+\alpha}\chi_p/\veee
{x}_L^{d+\alpha}$.
This completes the proof of (\ref{eq-subpcbd}), where
$K_p=\tilde\ell^{d+\alpha}\chi_p+C_p/(1-\varepsilon)$.
\end{pf*}

%s3 #&#
\section{Proof of the main result}\label{sproof}
In this section, we prove the asymptotic behavior~(\ref{eq-main}) of
$G_{p_{\mathrm{c}}}$
in high dimensions. To do so, we show in Section~\ref{ssopt-bound} that,
if $d>d_{\mathrm{c}}$ and $L\gg1$, then $G_p$ for $p\le p_{\mathrm
{c}}$ obeys
the same bound
as in
(\ref{eq-Spubd}) on $S_p$ for $p\le1$. Then, in Section~\ref{ssasymptotics},
we show that the obtained infrared bound on $G_{p_{\mathrm{c}}}$
implies its asymptotic
expression~(\ref{eq-main}). The proofs rely on the lace expansion~(\ref{eq-lace-exp}) for $G_p$.

%s3.1 #&#
\subsection{Bounds on \texorpdfstring{$\varPi_p$}{varPip} assuming the infrared bound on $G_p$}
In this subsection, we assume the infrared bound on $G_p$ and prove
bounds on
$\varPi_p$ and related quantities, such as its sum
$\hat\varPi_p(0)\equiv\sum_x\varPi_p(x)$, in high dimensions.
Before stating this more precisely, we need introduce the following parameter
for $\alpha>0$, $\alpha\ne2$ and \mbox{$d>\alpha\wedge2$} [cf., (\ref{eq-Spubd})]:
%
%e3.1 #&#
\begin{equation}
\label{eq-lambda} \lambda=\sup_{x\ne o}\frac{S_1(x)}{\veee{x}_L^{\alpha
\wedge
2-d}}=O
\bigl(L^{-\alpha
\wedge2} \bigr).
\end{equation}

%
%pr3.1 #&#
\begin{proposition}\label{propositiondiagbds}
Let $\alpha>0$, $\alpha\ne2$ and $d>d_{\mathrm{c}}$, and assume the
properties
(\ref{eq-D-def}) and (\ref{eq-hatDasy}) in Assumption~\ref{assumptiond}.
Suppose that
%
%e3.2 #&#
\begin{equation}
\label{eq-G-hyp} p\le3,\qquad G_p(x)\le3\lambda\veee{x}_L^{\alpha\wedge2-d}
\qquad[x\ne o].
\end{equation}
If $\lambda\ll1$ (i.e., $L\gg1$), then, for any $x\in\mathbb{Z}^d$,
%
%e3.3 #&#
%e3.4 #&#
\begin{eqnarray}
(pD*G_p) (x)&\le& O(\lambda) \veee{x}_L^{\alpha\wedge2-d},\label{eq-D*G-bd}
\\
\bigl|\varPi_p(x)-\delta_{o,x}\bigr|&\le& O \bigl(L^{-d}
\bigr)\delta_{o,x}+ O \bigl(\lambda^\ell \bigr)
\veee{x}_L^{(\alpha\wedge2-d)\ell}, \label{eq-varPi-bd}
\end{eqnarray}
where $\ell=2$ for percolation and $\ell=3$ for SAW and the Ising model.
As a result,
%
%e3.5 #&#
%e3.6 #&#
\begin{eqnarray}
\hat\varPi_p(0)&=&1+O \bigl(L^{-d} \bigr),\label{eq-varPi-sum}
\\
\bar\nabla^{\alpha\wedge2}\hat\varPi_p(0)&\equiv&\lim
_{|k|\to
0}\frac{\hat
\varPi_p(0)-\hat
\varPi_p
(k)}{1-\hat D(k)}
\nonumber\\[-8pt]\label{eq-bnabla-def} \\[-8pt]
&=& \cases{0, &\quad$[\alpha<2]$,
\vspace*{4pt}\cr
\displaystyle\frac{1}{\sigma^2}\sum
_x|x|^2\varPi_p(x)=O
\bigl(L^{-d(\ell-1)} \bigr), &\quad$[\alpha>2]$.}\nonumber
\end{eqnarray}
\end{proposition}

We prove this proposition by using the following lemma, which is an improved
version of \cite{hhs03}, Proposition~1.7.

%
%le3.2 #&#
\begin{lemma}\label{lemmaconv-bds}
\textup{(i)}~For any $a\ge b>0$ with $a+b>d$, there is an $L$-independent constant
$C=C(a,b,d)<\infty$ such that
%
%e3.7 #&#
\begin{eqnarray}
\label{eq-conv1} \sum_{y\in\mathbb{Z}^d}\veee{x-y}_L^{-a}
\veee{y}_L^{-b}\le\cases{ CL^{d-a}
\veee{x}_L^{-b}, &\quad$[a>d]$,
\vspace*{3pt}\cr
C
\veee{x}_L^{d-a-b},& \quad$[a<d]$.}
\end{eqnarray}

\textup{(ii)}
Let $f$ and $g$ be functions on $\mathbb{Z}^d$, with $g$ being
$\mathbb{Z}^d$-symmetric.
Suppose that there are $C_1,C_2,C_3>0$ and $\rho>0$ such that
%
%e3.8 #&#
\begin{equation}
f(x)=C_1\veee{x}_L^{\alpha\wedge2-d},\qquad \bigl|g(x)\bigr|\le
C_2\delta_{o,x}+C_3\veee{x}_L^{-d-\rho}.
\end{equation}
Then there is a $\rho'\in(0,\rho\wedge2)$ such that, for $d>\alpha
\wedge2$,
%
%e3.9 #&#
\begin{equation}
\label{eq-conv2} (f*g) (x)=\frac{C_1\|g\|_1}{\veee{x}_L^{d-\alpha\wedge
2}}+\frac{O(C_1C_3)} {
\veee{x}_L^{d-\alpha\wedge2+\rho'}}.
\end{equation}
\end{lemma}

\begin{pf*}{Proof of Proposition~\ref{propositiondiagbds}}
First, we note that
%
%e3.10 #&#
\begin{equation}
D(x)=\frac{O(L^\alpha)}{\veee{x}_L^{d+\alpha}}=\frac{O(L^\alpha)
\veee{x}_L^{-\alpha-\alpha\wedge2}}{\veee{x}_L^{d-\alpha\wedge
2}}\le\frac{O(\lambda)}{\veee{x}_L^{d-\alpha\wedge2}}.
\end{equation}
We also note that the identity $G_p(y)=\delta_{o,y}+G_p(y){\mathbh
1}_{\{y\ne o\}}$ holds
for all three models. Therefore, by using the assumed bound (\ref{eq-G-hyp}) and
Lemma~\ref{lemmaconv-bds}(i),\vadjust{\goodbreak} we obtain~(\ref{eq-D*G-bd}) as
%
%e3.11 #&#
\begin{eqnarray}
(D*G_p) (x)&=&D(x)+\sum_{y\ne o}D(x-y)
G_p(y)\nonumber
\\[-1pt]
&\le&\frac{O(\lambda)}{\veee{x}_L^{d-\alpha\wedge2}}+\sum_{y\in
\mathbb{Z}^d}
\frac{O
(L^\alpha)}{\veee{x-y}_L^{d+\alpha}} \frac{3\lambda}{\veee
{y}_L^{d-\alpha
\wedge2}}
\\[-1pt]
&\le&\frac{O(\lambda)}{\veee{x}_L^{d-\alpha\wedge2}}.
\nonumber
\end{eqnarray}

Inequality (\ref{eq-varPi-bd}) is obtained by repeatedly applying
(\ref{eq-G-hyp})--(\ref{eq-D*G-bd}) and Lem\-ma~\ref{lemmaconv-bds}(i) to the
diagrammatic bounds on $\varPi_p(x)$ in \cite{hhs03,s07} ($\varPi
_p(x)$ in
this paper equals $\delta_{o,x}+\Pi_z(x)$ in \cite{hhs03}, Proposition~1.8),
where $\ell$ is the number of disjoint paths in the diagrams from $o$
to $x$
(cf., Figure~\ref{figell}).
The proof is quite similar to \cite{hhs03}, Proposition~1.8 and
\cite{s07}, Proposition~3.1; the only difference is the use of
$\veee{\cdot}_L$ instead of $\veee{\cdot}_1$ and
Lemma~\ref{lemmaconv-bds}(i). Because of this, we gain the factor
$O(L^{-d}) (=O(\lambda)\veee{o}_L^{\alpha\wedge2-d})$ in (\ref{eq-varPi-bd}),
which is much smaller than $O(\lambda)$ as claimed in \cite{hhs03,s07}.

%
%f1 #&#
\begin{figure}%[b]

\includegraphics{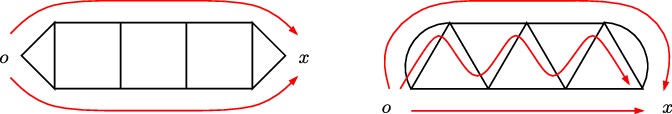}

%o\raisebox{-24pt}{\includegraphics[scale=.33]{perc}}x\qquad
%{}_{\raisebox{-23pt}{$o$}} \raisebox{-34pt}{\includegraphics[scale=.33]
%{sawIsing}} {}_{\raisebox{-23pt}{$x$}}
\caption{The left figure is an example of the lace-expansion diagrams
for percolation,
and the right one is for SAW and the Ising model. The number $\ell$ of
disjoint paths from $o$ to $x$ using different sets of line segments is
2 in the left figure and 3 in the right figure.}\label{figell}
\end{figure}

It remains to prove (\ref{eq-varPi-sum})--(\ref{eq-bnabla-def}).
By (\ref{eq-varPi-bd}), we readily obtain (\ref{eq-varPi-sum}) as
%
%e3.12 #&#
\begin{equation}
\qquad \hat\varPi_p(0)\equiv\sum_{x\in\mathbb{Z}^d}\varPi
_p(x)=1+O \bigl(L^{-d} \bigr)+O \bigl(L^{-d(\ell-1)}
\bigr) =1+O \bigl(L^{-d} \bigr).
\end{equation}
Moreover,
%
%e3.13 #&#
\begin{eqnarray}
&& \bigl|\hat\varPi_p(0)-\hat\varPi_p(k)\bigr|
\nonumber\\[-6pt]\\[-10pt]
&&\qquad \equiv\biggl|\sum
_{x\in\mathbb
{Z}^d} \bigl(1-\cos(k\cdot x) \bigr)
\varPi_p(x) \biggr|
\le O \bigl(\lambda^\ell \bigr)\sum_{x\in\mathbb{Z}^d}
\frac{1-\cos(k\cdot
x)}{\veee
{x}_L^{(d-\alpha
\wedge2)\ell}}.\nonumber
\end{eqnarray}
If $\alpha<2$, then there is a
$\delta\in(0,(2-\alpha)\wedge((\ell-1)(d-d_{\mathrm{c}})))$ such that
$1-\cos(k\cdot x)\le O(|k\cdot x|^{\alpha+\delta})$, hence
%
%e3.14 #&#
\begin{eqnarray}
\label{eq-varPi-diffalpha<2}
&& \bigl|\hat\varPi_p(0)-\hat\varPi_p(k)\bigr|\nonumber
\\
&&\qquad \le  O \bigl(|k|^{\alpha+\delta} \bigr)
\biggl(L^{-d\ell} \sum_{x\dvtx|x|\le L}|x|^{\alpha+\delta
}+L^{-\alpha\ell}
\sum_{x\dvtx|x|>L} \frac{|x|^{\alpha+\delta}}{|x|^{(d-\alpha)\ell}} \biggr)
\\
&&\qquad =O \bigl(L^{-d(\ell-1)+\alpha+\delta} \bigr)|k|^{\alpha+\delta}.\nonumber
\end{eqnarray}
If $\alpha>2$, then there is a $\delta\in(0,2\wedge((\ell
-1)(d-d_{\mathrm{c}})))$ such
that $1-\cos(k\cdot x)=\frac{1}2|k\cdot x|^2+O(|k\cdot x|^{2+\delta})$ and,
therefore,
%
%e3.15 #&#
\begin{eqnarray}
\label{eq-varPi-diffalpha>2}
&& \hat\varPi_p(0)-\hat\varPi_p(k)\nonumber
\\
&&\qquad = \frac{1}2\sum_{x\in\mathbb{Z}^d}|k\cdot x|^2\varPi_p(x)
+O \bigl(L^{-2\ell} \bigr)|k|^{2+\delta}\sum
_{x\in\mathbb{Z}^d}\frac
{|x|^{2+\delta}} {
\veee{x}_L^{(d-2)\ell}}
\\
&&\qquad =\frac{|k|^2}{2d}\sum_{x\in\mathbb{Z}^d}|x|^2
\varPi_p(x)+O \bigl(L^{-d(\ell
-1)+2+\delta} \bigr) |k|^{2+\delta}.\nonumber
\end{eqnarray}
Then, by the above estimates and (\ref{eq-hatDasy}), we obtain
%
%e3.16 #&#
\begin{eqnarray}
\label{eq-varPi-diff}
&& \frac{\hat\varPi_p(0)-\hat\varPi_p(k)}{1-\hat
D(k)}
\nonumber\\[-4pt]\\[-12pt]
&&\qquad = \cases{ O \bigl(L^{-d(\ell-1)+\delta}
\bigr)|k|^\delta, & \quad$[\alpha<2]$,
\vspace*{5pt}\cr
\displaystyle\frac{1}{\sigma^2} \sum_x|x|^2
\varPi_p(x)+O \bigl(L^{-d(\ell
-1)+\delta} \bigr)|k|^\delta, &
\quad$[\alpha>2]$,}\nonumber
\end{eqnarray}
hence (\ref{eq-bnabla-def}) by taking $|k|\to0$. This completes the
proof of
Proposition~\ref{propositiondiagbds}.
\end{pf*}

\begin{pf*}{Proof of Lemma~\ref{lemmaconv-bds}}
The proof of (\ref{eq-conv1}) is almost identical to that of
\cite{hhs03}, Proposition 1.7(i). However, since we are using $\veee
{\cdot}_L$
rather than $\veee{\cdot}_1$ as in \cite{hhs03}, we can gain the
extra factor
$L^{d-a}$ for $a>d$ in (\ref{eq-conv1}). To clarify this, we include
the proof here. First of all, since $a\ge b$, we have
%
%e3.17 #&#
\begin{eqnarray}
\label{eq-conv1-pr1}
\qquad&& \sum_{y\in\mathbb{Z}^d}\veee{x-y}_L^{-a}\veee{y}_L^{-b}\nonumber
\\
&&\qquad \le \sum_{y\dvtx|x-y|\le|y|}\veee{x-y}_L^{-a} \veee{y}_L^{-b}
+\sum_{y\dvtx|x-y|>|y|}\veee{x-y}_L^{-a}\veee{y}_L^{-b}
\\
&&\qquad \le 2\sum_{y\dvtx|x-y|\le|y|}\veee{x-y}_L^{-a}
\veee{y}_L^{-b}.\nonumber
\end{eqnarray}
Since $|x-y|\le|y|$ implies $|y|\ge\frac{1}2|x|$, we obtain that, for $a>d$,
%
%e3.18 #&#
\begin{eqnarray}
\nonumber && \sum_{y\dvtx|x-y|\le|y|}\veee{x-y}_L^{-a}
\veee{y}_L^{-b}
\\
&&\qquad \le 2^b\veee {x}_L^{-b} \sum_{y\in\mathbb{Z}^d}
\veee{x-y}_L^{-a}
\\
&&\qquad =\frac{C}2L^{d-a}\veee{x}_L^{-b}.\nonumber
\end{eqnarray}
For $a<d$, on the other hand, we use the identity
$1={\mathbh1}_{\{|y|\le(3/2)|x|\}}+{\mathbh1}_{\{|y|>(3/2)|x|\}
}$ and the fact that
$|y|>\frac{3}2|x|$ implies $|x-y|\ge\frac{1}3|y|$. Then we obtain
%
%e3.19 #&#
\begin{eqnarray}
\qquad && \sum_{y\dvtx|x-y|\le|y|}\veee{x-y}_L^{-a}\veee{y}_L^{-b}\nonumber
\\
&&\qquad \le 2^b\veee
{x}_L^{-b} \sum_{y\dvtx|x-y|\le|y|}
\veee{x-y}_L^{-a}{\mathbh1}_{\{|y|\le(3/2)|x|\}}\nonumber
\\
&&\quad\qquad{} +3^a\sum_{y\dvtx|x-y|\le|y|} \veee{y}_L^{-a-b}{\mathbh1}_{\{
|y|>(3/2)|x|\}}
\nonumber\\[-8pt]\\[-8pt]
&&\qquad \le 2^b\veee{x}_L^{-b}\sum
_{y\dvtx|x-y|\le(3/2)|x|}\veee{x-y}_L^{-a}\nonumber
\\
&&\quad\qquad{} +3^a\sum_{y\dvtx|y|>(3/2)|x|}
\veee{y}_L^{-a-b}\nonumber
\\
&&\qquad \le\frac{C}2\veee{x}_L^{d-a-b}.
\nonumber
\end{eqnarray}
This completes the proof of (\ref{eq-conv1}).

The proof of (\ref{eq-conv2}) is also quite similar to that of
\cite{hhs03}, Proposition 1.7(ii), where \cite{hhs03}, (5.8), is used.
However, \cite{hhs03}, (5.8), is valid only for $d>4$, not $d>2$ as claimed
in \cite{hhs03}, Proposition 1.7(ii). In fact, it is not difficult to avoid
this problem, and we include the proof here to clarify this. First, we note
that
%
%e3.20 #&#
\begin{equation}
\label{eq-f*geq} (f*g) (x)=\|g\|_1f(x)+\sum
_{y\in\mathbb{Z}^d}g(y) \bigl(f(x-y)-f(x) \bigr).
\end{equation}
To prove (\ref{eq-conv2}), it suffices to show that the sum in the right-hand
side is the error term in (\ref{eq-conv2}). For that, we split the sum into
the following three sums:
%
%e3.21 #&#
\begin{eqnarray}
\label{eq-splitting}
\sum_{y\in\mathbb{Z}^d}&=& \sum
_{y\dvtx|y|\le(1/3)|x|}+\sum_{y\dvtx|x-y|\le(1/3)|x|} +\sum
_{y\dvtx|y|\wedge|x-y|>(1/3)|x|}
\nonumber\\[-8pt]\\[-8pt]
&\equiv& {\sum_y}'+{\sum_y}''+{\sum_y}'''.\nonumber
\end{eqnarray}
It is not difficult to estimate the last two sums, as
%
%e3.22 #&#
\begin{eqnarray}
\label{eq-f*g}
\qquad && \biggl|{\sum_y}''g(y)\bigl(f(x-y)-f(x) \bigr) \biggr|\nonumber
\\
&&\qquad \le\frac{O(C_3)}{\veee{x}_L^{d+\rho}}\sum_{y\dvtx|x-y|\le(1/3)|x|}
\bigl(f(x-y) +f(x) \bigr)
\\
&&\qquad \le\frac{O(C_1C_3)}{\veee{x}_L^{d-\alpha\wedge2+\rho
}}\nonumber
\end{eqnarray}
and
%
%e3.23 #&#
\begin{eqnarray}
\label{eq-f*g} && \biggl|{\sum_y}'''g(y)
\bigl(f(x-y)-f(x) \bigr) \biggr|
\nonumber
\\
&&\qquad \le\frac{O(C_1)}{\veee{x}_L^{d-\alpha\wedge2}}\sum_{y\dvtx|y|>(1/3)|x|}
\frac{C_3}{\veee{y}_L^{d+\rho}}
\\
&&\qquad \le\frac{O(C_1C_3)}{\veee
{x}_L^{d-\alpha
\wedge2+\rho}}.
\nonumber
\end{eqnarray}
To estimate the sum $\sum_y'$, we use the $\mathbb{Z}^d$-symmetry of
$g$ to obtain
%
%e3.24 #&#
\begin{eqnarray}
&&{\sum_y}'g(y) \bigl(f(x-y)-f(x)
\bigr)
\nonumber\\[-8pt]\\[-8pt]
&&\qquad =\sum_{y\dvtx0<|y|\le (1/3)|x|}g(y) \biggl(
\frac
{f(x+y)+f(x-y)}2-f(x) \biggr).
\nonumber
\end{eqnarray}
Notice that
%
%e3.25 #&#
\begin{eqnarray}
\label{eq-HKbd-idea} && \biggl|\frac{f(x+y)+f(x-y)}2-f(x) \biggr|
\nonumber\\[-2pt]\\[-14pt]
&&\qquad \le\frac{O(C_1)}{\veee{x}_L^{d-\alpha\wedge2}}\times\cases{ 1, &\quad $\biggl[|x|\le
\dfrac{3}2L\biggr]$,
\vspace*{5pt}\cr
|y|^2/|x|^2, &\quad$
\biggl[|x|\ge\dfrac{3}2L\biggr]$.}
\nonumber
\end{eqnarray}
To verify this for $|x|\le\frac{3}2L$, we simply bound each $f$ by
$O(C_1)\veee{x}_L^{\alpha\wedge2-d}$. For $|x|\ge\frac{3}2L$, since
$|x\pm y|\ge|x|-|y|\ge\frac{2}3|x|\ge L$, we have
$f(x\pm y)=C_1|x\pm y|^{\alpha\wedge2-d}$. Then, by Taylor's theorem,
since $|{\pm}2\frac{x\cdot y}{|x|^2}+\frac{|y|^4}{|x|^4} |\le
\frac{7}9<1$,
we have
%
%e3.26 #&#
\begin{eqnarray}
|x\pm y|^{\alpha\wedge2-d}&=&|x|^{\alpha\wedge2-d} \biggl(1\pm2\frac
{x\cdot y} {
|x|^2}+
\frac{|y|^4}{|x|^4} \biggr)^{(\alpha\wedge2-d)/2}
\nonumber\\[-8pt]\\[-8pt]
&=&|x|^{\alpha\wedge2-d} \biggl(1\mp(d-\alpha\wedge2)\frac{x\cdot
y}{|x|^2} +O \biggl(
\frac{|y|^2}{|x|^2} \biggr) \biggr)
\nonumber
\end{eqnarray}
and (\ref{eq-HKbd-idea}) follows. Therefore, if $|x|\le\frac{3}2L$, then
$|y|\le\frac{1}2L$ and we obtain
%
%e3.27 #&#
\begin{eqnarray}
\label{eq-f*g|x|<2L} \qquad \biggl|{\sum_y}'g(y)
\bigl(f(x-y)-f(x) \bigr) \biggr|&\le&\frac
{O(C_1)}{\veee{x}_L^{d
-\alpha\wedge2}}\sum
_{y\dvtx0<|y|\le (1/2)L} \frac{C_3}{L^{d+\rho}}
\nonumber\\[-8pt]\\[-8pt]
&\le&\frac{O(C_1C_3)}{\veee{x}_L^{d-\alpha\wedge2+\rho}}.
\nonumber
\end{eqnarray}
If $|x|\ge\frac{3}2L$, then $\veee{x}_L=|x|$ and we obtain
%
%e3.28 #&#
\begin{eqnarray}
\label{eq-f*g|x|>2L} \qquad&& \biggl|{\sum_y}'g(y)
\bigl(f(x-y)-f(x) \bigr) \biggr|\nonumber
\\
&&\qquad \le\frac{O(C_1)}{\veee{x}_L^{d-\alpha\wedge2+2}}\sum_{y\dvtx0<|y|\le(1/3)|x|}
|y|^2 \biggl(\frac{C_3{\mathbh1}_{\{|y|\le L\}}}{L^{d+\rho}}+\frac
{C_3{\mathbh1}_{\{|y|>L\}}}{|y|^{d
+\rho}} \biggr)
\\
&&\qquad \le\frac{O(C_1C_3)}{\veee{x}_L^{d-\alpha\wedge2+2}}\times\cases{ L^{-\rho+2}, &\quad$[\rho>2]$,
\vspace*{2pt}\cr
\log|x|, &\quad$[\rho=2]$,
\vspace*{2pt}\cr
|x|^{2-\rho}, &\quad$[\rho<2]$.}
\nonumber
\end{eqnarray}
Summarizing the above yields (\ref{eq-conv2}). This completes the
proof of
Lemma~\ref{lemmaconv-bds}.
\end{pf*}

%s3.2 #&#
\subsection{Proof of the infrared bound on $G_p$}\label{ssopt-bound}
In this subsection, we prove that the hypothesis of
Proposition~\ref{propositiondiagbds} indeed holds for $p\le
p_{\mathrm{c}}$ in high
dimensions. The precise statement is the following.

%
%th3.3 #&#
\begin{theorem}\label{theoremIRbd}
Let $\alpha>0$, $\alpha\ne2$ and $d>d_{\mathrm{c}}$, and assume the
properties~(\ref{eq-D-def}), (\ref{eq-hatDasy}) and (\ref{eq-HK2}) in
Assumption~\ref{assumptiond}. Then, for $L\gg1$ and $p\le p_{\mathrm{c}}$,
%
%e3.29 #&#
\begin{equation}
G_p(x)\le O \bigl(L^{-\alpha\wedge2} \bigr)\veee{x}_L^{\alpha\wedge2-d}
\qquad[x\ne o].
\end{equation}
\end{theorem}

\begin{pf}
Let
%
%e3.30 #&#
\begin{equation}
\label{eq-g-def} g_p=p\vee\sup_{x\ne o}
\frac{G_p(x)}{\lambda\veee{x}_L^{\alpha
\wedge2-d}},
\end{equation}
where we recall the definition (\ref{eq-lambda}) of $\lambda$.
Suppose that
the following properties hold:
\begin{longlist}[(iii)]
\item[(i)]
$g_p$ is continuous (and nondecreasing) in $p\in[1,p_{\mathrm{c}})$.

\item[(ii)]
$g_1\le1$.

\item[(iii)]
If $\lambda\ll1$ (i.e., $L\gg1$), then $g_p\le3$ implies $g_p\le2$
for every
$p\in(1,p_{\mathrm{c}})$.
\end{longlist}
If the above properties hold, then in fact $g_p\le2$ for all
$p<p_{\mathrm{c}}$,
as long
as $d>d_{\mathrm{c}}$ and $\lambda\ll1$. In particular,
$G_p(x)\le2\lambda\veee{x}_L^{\alpha\wedge2-d}$ for all $x\ne o$ and
$p<p_{\mathrm{c}}$ $(\le2)$. By Lemma~\ref{lemmaGcont}, we can extend
this bound
up to
$p=p_{\mathrm{c}}$, hence the proof completed.

Now we verify those properties (i)--(iii).\vspace*{9pt}

\textit{Verification of} (i).
It suffices to show that, for every $p_0\in(1,p_{\mathrm{c}})$,\break 
$\sup_{x\ne o}G_p(x)/\veee{x}_L^{\alpha\wedge2-d}$ is continuous in
$p\in[1,p_0]$. By the monotonicity of~$G_p(x)$ in $p\le p_0$
and using Lemma~\ref{lemmasubpcbd}, we have
%
%e3.31 #&#
\begin{equation}
\frac{G_p(x)}{\veee{x}_L^{\alpha\wedge2-d}}\le\frac{G_{p_0}(x)} {
\veee{x}_L^{\alpha\wedge2-d}}\le\frac{K_{p_0}\veee
{x}_L^{-d-\alpha}} {
\veee{x}_L^{\alpha\wedge2-d}}=
\frac{K_{p_0}}{\veee{x}_L^{\alpha
+\alpha
\wedge2}}.
\end{equation}
On the other hand, for any $x_0\ne o$ with $D(x_0)>0$, there exists an
$R=R(p_0,x_0)<\infty$ such that, for all $|x|\ge R$,
%
%e3.32 #&#
\begin{equation}
\frac{K_{p_0}}{\veee{x}_L^{\alpha+\alpha\wedge2}}\le\frac{D(x_0)} {
\veee{x_0}_L^{\alpha\wedge2-d}}.
\end{equation}
Moreover, by using $p\ge1$ and the lower bound of the second inequality
in (\ref{eq-RWbds}), we have
%
%e3.33 #&#
\begin{equation}
\frac{D(x_0)}{\veee{x_0}_L^{\alpha\wedge2-d}}\le\frac{pD(x_0)} {
\veee{x_0}_L^{\alpha\wedge2-d}}\le\frac{G_p(x_0)}{\veee
{x_0}_L^{\alpha
\wedge2-d}}.
\end{equation}
As a result, for any $p\in[1,p_0]$, we obtain
%
%e3.34 #&#
\begin{equation}
\sup_{x\ne o}\frac{G_p(x)}{\veee{x}_L^{\alpha\wedge2-d}}=\frac
{G_p(x_0)} {
\veee{x_0}_L^{\alpha\wedge2-d}}\vee\max
_{x\dvtx0<|x|<R} \frac{G_p(x)}{\veee{x}_L^{\alpha\wedge2-d}}.
\end{equation}
Since $G_p(x)$ is continuous in $p$ (cf., Lemma~\ref{lemmaGcont}) and the
maximum of finitely many continuous functions is continuous, we can conclude
that $g_p$ is continuous in $p\in[1,p_0]$, as required.\vspace*{9pt}

\textit{Verification of} (ii).
By the first inequality in (\ref{eq-RWbds}) and the definition
(\ref{eq-lambda}) of~$\lambda$, we readily obtain
%
%e3.35 #&#
\begin{equation}
g_1=1\vee\sup_{x\ne o}\frac{G_1(x)}{\lambda\veee{x}_L^{\alpha
\wedge
2-d}}\le1 \vee
\sup_{x\ne o}\frac{S_1(x)}{\lambda\veee{x}_L^{\alpha\wedge2-d}}=1
\end{equation}
as required.\vspace*{9pt}

\textit{Verification of} (iii).
If $d>d_{\mathrm{c}}$, $\lambda\ll1$ and $g_p\le3$, then, by
Proposition~\ref{propositiondiagbds}, $\varPi_p$ satisfies
(\ref{eq-varPi-bd})--(\ref{eq-bnabla-def}) as well as (\ref{eq-varPi-diff}).
We use these estimates and the lace expansion to prove $g_p\le2$ as follows.

First, we recall (\ref{eq-lace-exp}) and (\ref{eq-RW-lace}):
%
%e3.36 #&#
\begin{equation}
\label{eq-lace-again} G_p=\varPi_p+\varPi_p*pD*G_p,
\qquad S_p=\delta+pD*S_p,
\end{equation}
or equivalently
%
%e3.37 #&#
\begin{equation}
\label{eq-recursion} \varPi_p=G_p*(\delta-
\varPi_p*pD),\qquad \delta=(\delta-pD)*S_p.
\end{equation}
Inspired by the similarity of the above identities, we approximate
$G_p$ to
$r\varPi_p*S_q$ with some constant $r\in(0,\infty)$ and the
parameter change
$q\in[0,1]$. Rewrite $G_p$ as follows:
%
%e3.38 #&#
\begin{eqnarray}
\label{eq-approx-by-S} \qquad G_p&=&r\varPi_p*S_q+G_p* \delta-r\varPi_p*S_q\nonumber
\\
&=&r\varPi_p*S_q+G_p*(\delta-qD)*S_q-rG_p*(\delta-\varPi_p*pD)*S_q
\\
&=&r\varPi_p*S_q+G_p*E_{p,q,r}*S_q,\nonumber\
\end{eqnarray}
where
%
%e3.39 #&#
\begin{equation}
\label{eq-E-def} E_{p,q,r}=\delta-qD-r(\delta-\varPi_p*pD).
\end{equation}
We choose $q,r$ to satisfy
%
%e3.40 #&#
\begin{equation}
\label{eq-bc} \hat E_{p,q,r}(0)=\bar\nabla^{\alpha\wedge2}\hat
E_{p,q,r}(0)=0,
\end{equation}
or equivalently
%
%e3.41 #&#
\begin{equation}
\label{eq-simultaneous} \cases{ 1-q-r \bigl(1-\hat\varPi_p(0)p
\bigr)=0,
\vspace*{4pt}\cr
-q+r \bigl(\hat\varPi_p(0)+\bar
\nabla^{\alpha\wedge
2}\hat\varPi_p(0) \bigr)p=0.}
\end{equation}
Solving these simultaneous equations for $r$ and using (\ref{eq-bnabla-def}),
we obtain
%
%e3.42 #&#
\begin{eqnarray}
\label{eq-r-sol} r= \bigl(1+p\bar\nabla^{\alpha\wedge2}\hat\varPi_p(0)
\bigr)^{-1}=1+ \cases{ 0, &\quad$[\alpha<2]$,
\vspace*{2pt}\cr
O
\bigl(L^{-d(\ell-1)} \bigr), &\quad$[\alpha>2]$.}
\end{eqnarray}
On the other hand, by taking the Fourier transform of (\ref{eq-lace-again})
and setting $k=0$, we obtain
%
%e3.43 #&#
\begin{equation}
\label{eq-chi-identity} \chi_p=\hat\varPi_p(0)+\hat
\varPi_p(0)p\chi_p,
\end{equation}
or equivalently $\hat\varPi_p(0)=\chi_p/(1+p\chi_p)$ and, therefore,
%
%e3.44 #&#
\begin{equation}
\label{eq-q-sol} q=1-r \bigl(1-\hat\varPi_p(0)p \bigr)=1-
\frac{r}{1+p\chi_p}\in(0,1],
\end{equation}
where we have used $p\ge1$, $\chi_p\ge1$ and (\ref{eq-r-sol}) to guarantee
the positivity (by taking $L\gg1$ if $\alpha>2$).

In addition, by solving (\ref{eq-chi-identity}) for $\chi_p$ and using
(\ref{eq-varPi-sum}), we have
%
%e3.45 #&#
\begin{equation}
\label{eq-chi-rep} \chi_p=\frac{\hat\varPi_p(0)}{1-\hat\varPi
_p(0)p}=\frac{1+O(L^{-d})} {
1-\hat\varPi_p(0)p},
\end{equation}
hence $1-\hat\varPi_p(0)p\ge0$. In particular,
$p\le\hat\varPi_p(0)^{-1}=1+O(L^{-d})\le2$, as required.

It remains to prove $G_p(x)\le2\lambda\veee{x}_L^{\alpha\wedge2-d}$.
To do so, we use the following property of $E_{p,q,r}$.

%
%pr3.4 #&#
\begin{proposition}\label{propositionEconv}
Let $q,r$ be defined as in (\ref{eq-r-sol})--(\ref{eq-q-sol}).
Under the hypothesis of Proposition~\ref{propositiondiagbds}, there is
a $\rho\in(0,\alpha\wedge2)$ such that
%
%e3.46 #&#
\begin{equation}
\label{eq-E*S-bd} \bigl|(E_{p,q,r}*S_q) (x)\bigr|\le O
\bigl(L^{-d(\ell-1)} \bigr) \biggl({\mathbh1}_{\{\alpha
>2\}}
\delta_{o,x} +\frac{L^\rho}{\veee{x}_L^{d+\rho}} \biggr).
\end{equation}
\end{proposition}

For now, we assume this proposition and complete verifying the property (iii).
First, by rearranging (\ref{eq-approx-by-S}) and using
\mbox{$S_q\le S_1$} as well as (\ref{eq-varPi-sum}) and (\ref{eq-r-sol}) for
\mbox{$L\gg 1$}, we
obtain
%
%e3.47 #&#
\begin{eqnarray}
\label{eq-bootstrapping-fin}
\qquad G_p&=&r\varPi_p*S_q+G_p*E_{p,q,r}*S_q\nonumber
\\
&=&r\hat\varPi_p(0)S_q-r \bigl(\hat
\varPi_p(0)\delta-\varPi_p \bigr)*S_q+G_p*
E_{p,q,r}*S_q
\\
&\le& \bigl(1+O \bigl(L^{-d} \bigr) \bigr)S_1-r \bigl(
\hat \varPi_p(0)\delta-\varPi_p \bigr)*
S_q+G_p*E_{p,q,r}*S_q.
\nonumber
\end{eqnarray}
Then, by Proposition~\ref{propositionEconv} and
Lemma~\ref{lemmaconv-bds}(i), the third term is bounded as
%
%e3.48 #&#
\begin{eqnarray}
&&\bigl|(G_p*E_{p,q,r}*S_q) (x)\bigr|\nonumber
\\
&&\qquad \le O \bigl(L^{-d(\ell-1)} \bigr)\sum_{y\in\mathbb{Z}^d}
\frac{3\lambda}{\veee{y}_L^{d-\alpha
\wedge2}} \biggl(\delta_{y,x}+\frac{L^\rho}{\veee{x-y}_L^{d+\rho
}} \biggr)
\\
&&\qquad \le\frac{O(L^{-d(\ell-1)})\lambda}{\veee{x}_L^{d-\alpha\wedge
2}}.\nonumber
\end{eqnarray}
Also, by (\ref{eq-varPi-bd}) and Lemma~\ref{lemmaconv-bds}(i), the second
term in (\ref{eq-bootstrapping-fin}) is bounded as
%
%e3.49 #&#
\begin{eqnarray}
&&\bigl| \bigl( \bigl(\hat\varPi_p(0)\delta-\varPi_p
\bigr)*S_q \bigr) (x) \bigr|\nonumber
\\
&&\qquad  = \biggl|\sum_{y\ne o}\varPi_p(y) \bigl(S_q(x)-S_q(x-y) \bigr) \biggr|
\nonumber\\[-8pt]\\[-8pt]
&&\qquad \le \sum_{y\ne o}\bigl|\varPi_p(y)\bigr|
S_q(x)+\sum_{y\ne o}\bigl|\varPi_p(y)\bigr|
S_q(x-y)\nonumber
\\
&&\qquad \le \frac{O(L^{-d(\ell-1)})\lambda}{\veee{x}_L^{d-\alpha\wedge 2}}.\nonumber
\end{eqnarray}
Putting these estimates back into (\ref{eq-bootstrapping-fin}), we
obtain that,
for $L\gg1$,
%
%e3.50 #&#
\begin{equation}
\qquad G_p(x)\le \bigl(1+O \bigl(L^{-d} \bigr) \bigr)\frac
\lambda{\veee{x}_L^{d-\alpha
\wedge2}} +\frac{O(L^{-d(\ell-1)})\lambda}{\veee{x}_L^{d-\alpha\wedge
2}}\le
\frac{2
\lambda}{\veee{x}_L^{d-\alpha\wedge2}}
\end{equation}
as required. This completes the proof of Theorem~\ref{theoremIRbd}
assuming Proposition~\ref{propositionEconv}.
\end{pf}

\begin{pf*}{Proof of Proposition~\ref{propositionEconv}}
First, by substituting $q=1-r(1-\hat\varPi_p(0)p)$ [cf., (\ref{eq-q-sol})] into
(\ref{eq-E-def}) and using $1-r=pr\bar\nabla^{\alpha\wedge2}\hat
\varPi
_p(0)$ [cf., (\ref{eq-r-sol})],
we obtain
%
%e3.51 #&#
\begin{equation}
\label{eq-E-rep} E_{p,q,r}=pr \bigl(\bar\nabla^{\alpha\wedge2}\hat\varPi
_p(0) (\delta-D)- \bigl(\hat\varPi_p(0) \delta-
\varPi_p \bigr)*D \bigr).
\end{equation}
Using this representation, we prove (\ref{eq-E*S-bd}) for $|x|\le2L$ and
$|x|>2L$, separately.

For $|x|\le2L$, we simply use (\ref{eq-Spubd}) to bound
$|(E_{p,q,r}*S_q)(x)|$ by
%
%e3.52 #&#
\begin{equation}
\label{eq-E*S-bd|x|<2L} \bigl|(E_{p,q,r}*S_q) (x)\bigr|\le\bigl|E_{p,q,r}(x)\bigr|+O
\bigl(L^{-d} \bigr)\sum_{y\in\mathbb
{Z}^d}\bigl|E_{p,q,r}(y)\bigr|.
\end{equation}
By (\ref{eq-E-rep}), we have
%
%e3.53 #&#
\begin{eqnarray}
\label{eq-E-bd} \qquad &&\frac{|E_{p,q,r}(x)|}{pr}\nonumber
\\
&&\qquad \le\bigl|\bar\nabla^{\alpha\wedge2}\hat\varPi_p(0)\bigr| \bigl(
\delta_{o,x}+D(x) \bigr)+ \bigl| \bigl( \bigl(\hat\varPi_p(0)
\delta-\varPi_p \bigr)*D \bigr) (x) \bigr|
\nonumber\\[-8pt]\\[-8pt]
&&\qquad =\bigl|\bar\nabla^{\alpha\wedge2}\hat\varPi_p(0)\bigr| \bigl(\delta
_{o,x}+D(x) \bigr)+ \biggl|\sum_{z\ne o}
\varPi_p(z) \bigl(D(x)-D(x-z) \bigr) \biggr|
\nonumber
\\
&&\qquad \le\bigl|\bar\nabla^{\alpha\wedge2}\hat\varPi_p(0)\bigr| \bigl(
\delta _{o,x}+D(x) \bigr)+\sum_{z\ne
o}\bigl|
\varPi_p (z)\bigr| \bigl(D(x)+D(x-z) \bigr).
\nonumber
\end{eqnarray}
Using (\ref{eq-varPi-bd})--(\ref{eq-bnabla-def}) and (\ref{eq-r-sol}), we obtain that
%
%e3.54 #&#
\begin{eqnarray}
&& \bigl|E_{p,q,r}(x)\bigr|\nonumber
\\
&&\qquad \le  O \bigl(L^{-d(\ell-1)} \bigr){
\mathbh1}_{\{\alpha>2\}} \bigl( \delta_{o,x}+O \bigl(L^{-d}
\bigr) \bigr)+O \bigl(L^{-d} \bigr)\sum_{z\ne o}\bigl|
\varPi_p(z)\bigr|
\\
&&\qquad \le  O \bigl(L^{-d(\ell-1)} \bigr){\mathbh1}_{\{\alpha>2\}}\delta
_{o,x}+O \bigl(L^{-d\ell} \bigr)\nonumber
\end{eqnarray}
and that, by summing (\ref{eq-E-bd}) over $x\in\mathbb{Z}^d$,
%
%e3.55 #&#
\begin{eqnarray}
&&O \bigl(L^{-d} \bigr)\sum_{x\in\mathbb{Z}^d}\bigl|E_{p,q,r}(x)\bigr|\nonumber
\\
&&\qquad \le O \bigl(L^{-d} \bigr) \biggl(2\bigl|\bar\nabla^{\alpha\wedge2}
\hat \varPi_p (0)\bigr|+2\sum_{z\ne o}\bigl|
\varPi_p(z)\bigr| \biggr)
\\
&&\qquad \le O \bigl(L^{-d\ell} \bigr).\nonumber
\end{eqnarray}
Therefore, for $|x|\le2L$,
%
%e3.56 #&#
\begin{eqnarray}
\bigl|(E_{p,q,r}*S_q) (x)\bigr|&\le& O \bigl(L^{-d(\ell-1)}
\bigr){ \mathbh1}_{\{\alpha>2\}
}\delta_{o,x}+O \bigl(L^{-d
\ell}\bigr)
\nonumber\\[-8pt]\\[-8pt]
&\le& O \bigl(L^{-d(\ell-1)} \bigr){\mathbh1}_{\{\alpha>2\}}
\delta_{o,x}+\frac
{O(L^{-d(\ell-1)+
\rho})}{\veee{x}_L^{d+\rho}}.
\nonumber
\end{eqnarray}

It remains to prove (\ref{eq-E*S-bd}) for $|x|>2L$. To do so, we first rewrite
$(E_{p,q,r}*S_q)(x)$ as
%
%e3.57 #&#
\begin{eqnarray}\label{eq-E*S-Fourier}
\qquad (E_{p,q,r}*S_q) (x)&=&\int
_{[-\pi,\pi]^d}\frac{\mathrm{d}^dk}{(2\pi)^d} \hat{E}_{p,q,r}(k)
\frac{e^{-ik\cdot x}}{1-q\hat{D}(k)}
\nonumber\\[-8pt]\\[-8pt]
&=&\int_0^\infty\mathrm{d}t\int
_{[-\pi,\pi]^d}\frac{\mathrm
{d}^dk}{(2\pi)^d} \hat{E}_{p,q,r}(k)
e^{-t(1-q\hat{D}(k))-ik\cdot x}.
\nonumber
\end{eqnarray}
Then we split the integral with respect to $t$ into $\int_0^T$ and
$\int_T^\infty$, where $T$ is arbitrary for now, but it will be determined
shortly. For the latter integral, we use the Fourier transform of
(\ref{eq-E-rep}), which is
%
%e3.58 #&#
\begin{equation}
\label{eq-E-Fourier}
\qquad \hat E_{p,q,r}(k)=pr \bigl(1-\hat D(k) \bigr) \biggl(\bar
\nabla^{\alpha
\wedge
2}\hat\varPi_p(0) -\frac{\hat\varPi_p(0)-\hat\varPi_p(k)}{1-\hat
D(k)}\hat D(k)
\biggr).
\end{equation}
Because of (\ref{eq-hatDasy}), (\ref{eq-bnabla-def}) and (\ref{eq-varPi-diff}),
there is a $\delta>0$ such that
%
%e3.59 #&#
\begin{equation}
\label{eq-E-Fourier|k|<<1} \hat E_{p,q,r}(k)=O \bigl(L^{-d(\ell-1)+\alpha
\wedge2+\delta}
\bigr)|k|^{\alpha
\wedge2
+\delta}\qquad[\alpha\ne2].
\end{equation}
Since $1-q\hat{D}(k)\ge q(1-\hat{D}(k))$, the contribution to
(\ref{eq-E*S-Fourier}) from the large-$t$ integral is bounded as
%
%e3.60 #&#
\begin{eqnarray}
&& \biggl|\int_T^\infty\mathrm{d}t\int
_{[-\pi,\pi]^d}\frac{\mathrm
{d}^dk}{(2\pi)^d} \hat{E}_{p,q,r}(k)
e^{-t(1-q\hat{D}(k))-ik\cdot x} \biggr|
\nonumber\\[-8pt]\\[-8pt]
&&\qquad \le O \bigl(L^{-d(\ell-1)+\alpha\wedge2+\delta} \bigr)\int_T^\infty\mathrm{d}t \int_{[-\pi,\pi]^d}\frac{\mathrm{d}^dk}{(2\pi)^d}|k|^{\alpha
\wedge 2+\delta} e^{-tq(1-\hat{D}(k))}.\hspace*{-30pt}\nonumber
\end{eqnarray}
Since $p\ge1$, we have $q\ge1-r/(1+\chi_1)\ge1-r/2$ [cf., (\ref{eq-q-sol})],
which is bounded away from zero when $L\gg1$. Therefore, by using
(\ref{eq-hatDasy}), we obtain
%
%e3.61 #&#
\begin{equation}
\qquad \int_{[-\pi,\pi]^d}\frac{\mathrm{d}^dk}{(2\pi)^d}|k|^{\alpha
\wedge 2+\delta}
e^{-tq(1-\hat{D}(k))} =O \bigl(L^{\alpha\wedge2}t \bigr)^{-1-((d+\delta)/(\alpha\wedge2))},
\end{equation}
hence
%
%e3.62 #&#
\begin{eqnarray}
\label{eq-E*Slarge-t}
&& \biggl|\int_T^\infty\mathrm{d}t\int
_{[-\pi,\pi]^d}\frac{\mathrm
{d}^dk}{(2\pi)^d} \hat{E}_{p,q,r}(k)
e^{-t(1-q\hat{D}(k))-ik\cdot x} \biggr|
\nonumber\\[-8pt]\\[-8pt]
&&\qquad \le O \bigl(L^{-d\ell} \bigr) T^{-(d+\delta)/(\alpha\wedge2)}.\nonumber
\end{eqnarray}
Let
%
%e3.63 #&#
\begin{equation}
\label{eq-rhoT} \rho=\frac{(\alpha\wedge2)\delta}{d+\alpha\wedge2+\delta
},\qquad T= \biggl(\frac{|x|}L
\biggr)^{\alpha\wedge2-\rho}.
\end{equation}
Then, since $|x|>2L$,
%
%e3.64 #&#
\begin{equation}
O \bigl(L^{-d\ell} \bigr)T^{-(d+\delta)/(\alpha\wedge2)}=\frac
{O(L^{-d(\ell
-1)+\rho})} {
\veee{x}_L^{d+\rho}}.
\end{equation}

To estimate the contribution to (\ref{eq-E*S-Fourier}) from the small-$t$
integral, we use the identity
%
%e3.65 #&#
\begin{eqnarray}
\label{eq-E*Ssmall-t} &&\int_0^T\mathrm{d}t\int
_{[-\pi,\pi]^d}\frac{\mathrm{d}^dk}{(2\pi)^d} \hat{E}_{p,q,r}(k)
e^{-t(1-q\hat{D}(k))-ik\cdot x}
\nonumber\\[-8pt]\\[-8pt]
&&\qquad =\int_0^T\mathrm{d}t\,e^{-t}
\sum_{n=0}^\infty\frac{(tq)^n}{n!}
\bigl(E_{p,q,r}* D^{*n} \bigr) (x),\nonumber
\end{eqnarray}
where, by (\ref{eq-E-rep}) and (\ref{eq-bnabla-def}),
%
%e3.66 #&#
\begin{eqnarray}
\label{eq-E*D*n-rep} \bigl(E_{p,q,r}*D^{*n} \bigr) (x)
&=& \underbrace{pr\bar\nabla^{\alpha\wedge2}\hat\varPi _p(0)}_{O(L^{-d(\ell
-1)}){\mathbh1}_{\{\alpha>2\}}}
\sum_{y\in\mathbb{Z}^d}D(y) \bigl(D^{*n}(x)-D^{*n}(x-y)
\bigr)
\nonumber\\[-8pt]\\[-8pt]
&&{} -pr\sum_{y\in\mathbb{Z}^d}\varPi_p(y)
\bigl(D^{*(n+1)}(x)-D^{*(n+1)}(x-y) \bigr).\nonumber
\end{eqnarray}
In the following, we use the decomposition (\ref{eq-splitting}) of
$\sum_y$ and
estimate the contribution to (\ref{eq-E*Ssmall-t}) from $\sum_y'$,
$\sum_y''$
and $\sum_y'''$, separately.

First, we estimate the contribution from
$\sum_y''\equiv\sum_{y\dvtx|x-y|\le (1/3)|x|}$. Since
$|y|\ge|x|-|x-y|\ge\frac{2}3|x|$ in this domain of summation, we bound
$|\varPi_p(y)|$ by $O(\lambda^\ell)\veee{x}_L^{(\alpha\wedge
2-d)\ell}$ [cf., (\ref{eq-varPi-bd})] and then use (\ref{eq-HK1}), $\sum_y''1\le
O(\veee{x}_L^d)$ and $\sum_y''D^{*(n+1)}(x-y)\le1$. As a result,
%
%e3.67 #&#
\begin{eqnarray}
&& \biggl|{\sum_y}''
\varPi_p(y) \bigl(D^{*(n+1)}(x)-D^{*(n+1)}(x-y) \bigr)
\biggr|\nonumber
\\[-4pt]
&&\qquad \le\frac{O(\lambda^\ell)}{\veee{x}_L^{(d-\alpha\wedge2)\ell
}} \biggl(\frac{O
(L^{\alpha\wedge2}) n}{\veee{x}_L^{\alpha\wedge2}}+1 \biggr)
\\[-1pt]
&&\qquad \le\frac{O(L^{-d(\ell-1)+\alpha\wedge2})}{\veee{x}_L^{d+\alpha
\wedge2}} \biggl(\frac{O(L^{\alpha\wedge2}) n}{\veee{x}_L^{\alpha\wedge
2}}+1 \biggr).
\nonumber
\end{eqnarray}
Similarly, for $\alpha>2$,
%
%e3.68 #&#
\begin{eqnarray}
&&O \bigl(L^{-d(\ell-1)} \bigr) \biggl|{\sum_y}''D(y)
\bigl(D^{*n}(x)-D^{*n}(x-y) \bigr) \biggr|\nonumber
\\[-3pt]
&&\qquad \le\frac{O(L^{-d(\ell-1)+\alpha})}{\veee{x}_L^{d+\alpha}} \biggl(\frac{O
(L^2) n}{\veee{x}_L^2}+1 \biggr)
\\[-1pt]
&&\qquad \le\frac{O(L^{-d(\ell-1)+2})}{\veee{x}_L^{d+2}} \biggl(\frac
{O(L^2) n} {
\veee{x}_L^2}+1 \biggr).
\nonumber
\end{eqnarray}

To\vspace*{1pt} estimate the contribution to (\ref{eq-E*Ssmall-t}) from
$\sum_y'''\equiv\sum_{y\dvtx|y|\wedge|x-y|>(1/3)|x|}$ in~(\ref{eq-E*D*n-rep}),
we bound $D^{*(n+1)}(x)$ and $D^{*(n+1)}(x-y)$ by
$O(L^{\alpha\wedge2})n/\veee{x}_L^{d+\alpha\wedge2}$ and then use
(\ref{eq-varPi-bd}) to bound $|\varPi_p(y)|$. The result is
%
%e3.69 #&#
\begin{eqnarray}
&& \biggl|{\sum_y}'''
\varPi_p(y) \bigl(D^{*(n+1)}(x)-D^{*(n+1)}(x-y)
\bigr) \biggr|\nonumber
\\[-4pt]
&&\qquad \le\frac{O(L^{\alpha\wedge2}) n}{\veee{x}_L^{d+\alpha\wedge
2}}\sum_{y\dvtx|y|
>(1/3)|x|}\bigl|\varPi_p(y)\bigr|
\\[-1pt]
&&\qquad \le\frac{O(L^{-(\ell-1)(\alpha\wedge2)}) n}{\veee
{x}_L^{d+2(\alpha
\wedge2)
+(\ell-1)(d-d_{\mathrm{c}})}} \le\frac{O(L^{-d(\ell-1)+2(\alpha\wedge
2)}) n}{\veee
{x}_L^{d+2(\alpha
\wedge
2)}}.
\nonumber
\end{eqnarray}
Similarly, for $\alpha>2$,
%
%e3.70 #&#
\begin{eqnarray}
&&O \bigl(L^{-d(\ell-1)} \bigr) \biggl|{\sum_y}'''D(y)
\bigl(D^{*n}(x)-D^{*n}(x-y) \bigr) \biggr|\nonumber
\\
&&\qquad \le\frac{O(L^{-d(\ell-1)+2}) n}{\veee{x}_L^{d+2}}\sum_{y\dvtx|y|>(1/3)|x|}D(y)
\\
&&\qquad \le
\frac{O(L^{-d(\ell-1)+4}) n}{\veee{x}_L^{d+4}}.\nonumber
\end{eqnarray}

Finally, we estimate the contribution to (\ref{eq-E*Ssmall-t}) from
$\sum_y'\equiv\sum_{y\dvtx|y|\le(1/3)|x|}$ in (\ref{eq-E*D*n-rep}).
By the
$\mathbb{Z}^d$-symmetry of $\varPi_p$ and using (\ref{eq-varPi-bd}) and
the assumption (\ref{eq-HK2}), we obtain
%
%e3.71 #&#
\begin{eqnarray}\label{eq-sumvarPi-bd}
\qquad\quad && \biggl|{\sum_y}'\varPi_p(y) \bigl(D^{*(n+1)}(x)-D^{*(n+1)}(x-y) \bigr)\biggr|\nonumber
\\
&&\qquad = \biggl|{\sum_y}'
\varPi_p(y) \biggl(D^{*(n+1)}(x)-\frac{D^{*(n+1)}(x+y)
+D^{*(n+1)}(x-y)}2 \biggr)
\biggr|\nonumber
\\
&&\qquad \le\frac{O(L^{\alpha\wedge2}) n}{\veee{x}_L^{d+\alpha\wedge
2+2}}\sum_{y\dvtx
|y|\le (1/3)|x|}
\frac{O(\lambda^\ell)\veee{y}_L^2}{\veee
{y}_L^{(d-\alpha
\wedge2)\ell}}
\nonumber\\[-8pt]\\[-8pt]\nonumber
&&\qquad \le\frac{O(L^{-(\ell-1)(\alpha\wedge2)}) n}{\veee{x}_L^{d+\alpha
\wedge2 +2}}\nonumber
\\
&&\quad\qquad{} \times\cases{ \veee{x}_L^{d+2-(d-\alpha\wedge2)\ell},
&\quad$ \bigl[d+2>(d-\alpha\wedge2)\ell \bigr]$,
\vspace*{4pt}\cr
1+\log\bigl(\veee{x}_L/L\bigr), &\quad$ \bigl[d+2=(d-\alpha\wedge2)\ell \bigr]$,
\vspace*{4pt}\cr
 L^{d+2-(d-\alpha\wedge2)\ell}, &\quad$ \bigl[d+2<(d-\alpha \wedge2)\ell
\bigr]$,}
\nonumber
\\
&&\qquad \le\frac{O(L^{-d(\ell-1)+2(\alpha\wedge2)}) n}{\veee
{x}_L^{d+2(\alpha
\wedge2)}},
\nonumber
\end{eqnarray}
where, to obtain the last inequality for $d+2=(d-\alpha\wedge2)\ell$,
which implies $\alpha<2$, we have used fact that
$(\veee{x}_L/L)^{\alpha-2}(1+\log(\veee{x}_L/L))$ is bounded.
Similarly, for $\alpha>2$,
%
%e3.72 #&#
\begin{eqnarray}
\label{eq-sumD-bd} \qquad\quad &&O \bigl(L^{-d(\ell-1)} \bigr) \biggl|{\sum
_y}'D(y) \bigl(D^{*n}(x)-D^{*n}(x-y)
\bigr) \biggr|\nonumber
\\
&&\qquad=O \bigl(L^{-d(\ell-1)} \bigr) \biggl|{\sum_y}'D(y)
\biggl(D^{*n}(x)-\frac{D^{*n}(x+y)+
D^{*n}(x-y)}2 \biggr) \biggr|
\\
&&\qquad \le\frac{O(L^{-d(\ell-1)+2}) n}{\veee{x}_L^{d+4}}\underbrace{\sum_{y\dvtx|y|
\le (1/3)|x|}
\frac{O(L^\alpha)\veee{y}_L^2}{\veee{y}_L^{d+\alpha
}}}_{O(L^2)}\nonumber
\\
&&\qquad  =\frac{O(L^{-d(\ell-1)+4}) n}{\veee{x}_L^{d+4}}. \nonumber
\end{eqnarray}

Now, by putting these estimates back into (\ref{eq-E*D*n-rep}), we obtain
%
%e3.73 #&#
\begin{equation}
\bigl| \bigl(E_{p,q,r}*D^{*n} \bigr) (x)\bigr|\le\frac{O(L^{-d(\ell-1)+\alpha\wedge
2})} {
\veee{x}_L^{d+\alpha\wedge2}}
\biggl(\frac{O(L^{\alpha\wedge2})} {
\veee{x}_L^{\alpha\wedge2}}n+1 \biggr),
\end{equation}
hence, by (\ref{eq-rhoT}),
%
%e3.74 #&#
\begin{eqnarray}
&& \Biggl|\int_0^T\mathrm{d}t\,e^{-t}\sum
_{n=0}^\infty\frac{(tq)^n}{n!}
\bigl(E_{p,q,r} *D^{*n} \bigr) (x) \Biggr|\nonumber
\\
&&\qquad \le\frac{O(L^{-d(\ell-1)+\alpha\wedge2})}{\veee{x}_L^{d+
\alpha\wedge2}} \biggl(\frac{O(L^{\alpha\wedge2})} {
\veee{x}_L^{\alpha\wedge2}}T^2+T
\biggr)
\\
&&\qquad \le\frac{O(L^{-d(\ell-1)+\alpha\wedge2})}{\veee{x}_L^{d+\alpha
\wedge2}}T =\frac{O(L^{-d(\ell-1)+\rho})}{\veee{x}_L^{d+\rho}}.
\nonumber
\end{eqnarray}
This completes the proof of Proposition~\ref{propositionEconv}.
\end{pf*}

%s3.3 #&#
\subsection{Derivation of the asymptotics of \texorpdfstring{$G_{p_{\mathrm{c}}}$}{Gpc}}\label{ssasymptotics}
Finally, we derive the asymptotic expression (\ref{eq-main}) for
$G_{p_{\mathrm{c}}}$.
First, by repeatedly applying (\ref{eq-approx-by-S}), we obtain
%
%e3.75 #&#
\begin{eqnarray}
G_p&=&r\varPi_p*S_q+G_p*E_{p,q,r}*S_q\nonumber
\\
&=&r\varPi_p*S_q+(r\varPi_p*S_q+G_p*E_{p,q,r}*S_q)*E_{p,q,r}*S_q\nonumber
\\
&=&r\varPi_p*S_q*(\delta+E_{p,q,r}*S_q)+G_p*(E_{p,q,r}*S_q)^{*2}
\\
& \vdots&\nonumber
\\
&=&r\varPi_p*S_q*\sum_{n=0}^{N-1}(E_{p,q,r}*S_q)^{*n}+G_p*(E_{p,q,r}*S_q)^{*N}.\nonumber
\end{eqnarray}
By Proposition~\ref{propositionEconv} and Lemma~\ref{lemmaconv-bds}(i),
we have that, for $p\le p_{\mathrm{c}}$,
%
%e3.76 #&#
\begin{equation}
\label{eq-E*S*n-bd} \bigl|(E_{p,q,r}*S_q)^{*n}(x)\bigr|\le O
\bigl(L^{-d(\ell-1)n} \bigr) \biggl({\mathbh1}_{\{
\alpha>2\}}
\delta_{o,x} +\frac{L^\rho}{\veee{x}_L^{d+\rho}} \biggr),
\end{equation}
hence, for any $N\in\mathbb{N}$,
%
%e3.77 #&#
\begin{equation}
\label{eq-E*S*sum-bd} \quad\qquad \sum_{n=0}^{N-1}\bigl|(E_{p,q,r}*S_q)^{*n}(x)\bigr|
\le \bigl(1+O \bigl(L^{-d(\ell
-1)} \bigr) \bigr) \delta_{o,x}+
\frac{O(L^{-d(\ell-1)+\rho})}{\veee{x}_L^{d+\rho}}.
\end{equation}
Therefore, we can take $N\to\infty$ to obtain that, for $p\le
p_{\mathrm{c}}$,
%
%e3.78 #&#
\begin{equation}
G_p=r\varPi_p*S_q*\sum
_{n=0}^\infty(E_{p,q,r}*S_q)^{*n}
\equiv H_p*S_q,
\end{equation}
where, by (\ref{eq-varPi-bd}) and (\ref{eq-E*S*sum-bd}),
%
%e3.79 #&#
\begin{eqnarray}
H_p(x)&=&r \Biggl(\varPi_p*\sum
_{n=0}^\infty(E_{p,q,r}*S_q)^{*n}
\Biggr) (x)\nonumber
\\
&=&r\sum_{y\in\mathbb{Z}^d} \biggl( \bigl(1+O
\bigl(L^{-d} \bigr) \bigr)\delta_{o,y}+\frac
{O(L^{-(\alpha
\wedge2)\ell})}{\veee{y}_L^{(d-\alpha\wedge2)\ell}}
\biggr)
\\
&&\hspace*{25pt}{} \times \biggl( \bigl(1+O \bigl(L^{-d(\ell-1)} \bigr) \bigr)
\delta_{y,x}+\frac{O(L^{-d(\ell-1)+\rho})}{\veee{x-y}_L^{d+\rho
}} \biggr).\nonumber
\end{eqnarray}
Notice that, by Lemma~\ref{lemmaconv-bds}(i) and using (\ref{eq-r-sol}) and
$d+\rho<(d-\alpha\wedge2)\ell$,
%
%e3.80 #&#
\begin{equation}
\label{eq-H-def} H_p(x)= \bigl(r+ O \bigl(L^{-d} \bigr)
\bigr)\delta_{o,x}+\frac{O(L^{-d(\ell-1)+\rho})}{\veee
{x}_L^{d+\rho}}.
\end{equation}

Now we set $p=p_{\mathrm{c}}$, so, by (\ref{eq-q-sol}), $q=1$.
By Proposition~\ref{propositionS} and Lem\-ma~\ref{lemmaconv-bds}(ii),
we obtain the asymptotic expression
%
%e3.81 #&#
\begin{equation}
\qquad G_{p_{\mathrm{c}}}(x)=\hat H_{p_{\mathrm{c}}}(0)\frac{\gamma_\alpha
/v_\alpha}{\veee
{x}_L^{d-\alpha
\wedge2}}+
\frac{O(L^{-\alpha\wedge2+\mu})}{|x|^{d-\alpha\wedge
2+\mu
}}+\frac{O
(L^{-d(\ell-1)-\alpha\wedge2+\rho})}{\veee{x}_L^{d-\alpha\wedge
2+\rho'}}.
\end{equation}
Since $H_{p_{\mathrm{c}}}$ is absolutely summable, we can change the
order of the limit
and the sum as
%
%e3.82 #&#
\begin{eqnarray}
\hat H_{p_{\mathrm{c}}}(0)&=&\lim_{|k|\to0}\hat
H_{p_{\mathrm
{c}}}(k)\nonumber
\\
&=&\lim_{|k|\to0}r\hat\varPi_{p_{\mathrm{c}}}(k)\sum
_{n=0}^\infty \bigl(\hat E_{p_{\mathrm{c}},1,r}(k) \hat
S_1(k) \bigr)^n
\\
&=&r\hat\varPi_{p_{\mathrm{c}}}(0)+r\hat\varPi_{p_{\mathrm
{c}}}(0)\sum
_{n=1}^\infty \Bigl(\lim_{|k|\to0} \hat
E_{p_{\mathrm{c}},1,r}(k) \hat S_1(k) \Bigr)^n.
\nonumber
\end{eqnarray}
By (\ref{eq-chi-rep}) and the fact that $\chi_p$ diverges as
$p\uparrow p_{\mathrm{c}}$,
we have $\hat\varPi_{p_{\mathrm{c}}}(0)=p_{\mathrm{c}}^{-1}$.
Moreover, by (\ref{eq-E-Fourier}) and
(\ref{eq-varPi-diff}),
%
%e3.83 #&#
\begin{equation}
\hat E_{p_{\mathrm{c}},1,r}(k) \hat S_1(k)=\hat E_{p_{\mathrm
{c}},1,r}(k)
\bigl(1-\hat D(k) \bigr)^{-1}\underset{|k|\to0}\to0.
\end{equation}
Therefore,
%
%e3.84 #&#
\begin{equation}
A=\hat H_{p_{\mathrm{c}}}(0)^{-1}=\frac{p_{\mathrm{c}}}r\equiv
p_{\mathrm{c}} \bigl(1+p_{\mathrm{c}}\bar\nabla^{\alpha
\wedge2}\hat\varPi
_{p_{\mathrm{c}}}(0) \bigr).
\end{equation}
This completes the proof of Theorem~\ref{theoremmain}.

\setcounter{equation}{0}
\setcounter{theorem}{0}
\begin{appendix}
%s4 #&#
\section*{Appendix: Verification of Assumption~\texorpdfstring{\lowercase{\protect\ref{assumptiond}}}{1.1}}\label{aeg123}
In this appendix, we show that the $\mathbb{Z}^d$-symmetric 1-step distribution
$D$ in (\ref{eqeg123}), defined more precisely below, satisfies the properties
(\ref{eq-D-def}), (\ref{eq-hatDasy}), (\ref{eq-Dbd}), (\ref{eq-HK1}) and (\ref{eq-HK2}) in
Assumption~\ref{assumptiond}.

First, for $\alpha>0$ and $\alpha\ne2$, we define
%
%e4.1 #&#
\begin{equation}
T_\alpha(t)=\frac{t^{-1-\alpha/2}}{\sum_{s\in\mathbb
{N}}s^{-1-\alpha/2}} \qquad[t\in\mathbb{N}].
\end{equation}
Next, let $h$ be a nonnegative bounded function on $\mathbb{R}^d$ that
is piecewise\break 
continuous, $\mathbb{Z}^d$-symmetric, supported in $[-1,1]^d$ and normalized
[i.e.,\break $\int_{[-1,1]^d}h(x) \,\mathrm{d}^dx=1$]; for example,
$h(x)=2^{-d}{\mathbh1}_{\{\|x\|_\infty\le1\}}$. Then, for large $L$
(to ensure
positivity of the denominator), we define
%
%e4.2 #&#
\begin{equation}
U_L(x)=\frac{h(x/L)}{\sum_{y\in\mathbb{Z}^d}h(y/L)}\qquad \bigl[x\in \mathbb{Z}^d
\bigr],
\end{equation}
where (cf., \cite{csIII,hs02})
%
%e4.3 #&#
%e4.4 #&#
\begin{eqnarray}
&& \sigma_L^2 \equiv \sum_{x\in\mathbb{Z}^d}|x|^2U_L(x)=O
\bigl(L^2 \bigr), \label{eq-sigmaL2}
\\
&& \hat U_L(k) \cases{\displaystyle =1-\frac{\sigma_L^2}{2d}|k|^2+O
\bigl(\bigl(L|k|\bigr)^{2+\zeta
} \bigr), &\quad$\bigl[|k|\to0\bigr]$,
\vspace*{2pt}\cr
\in(-1+
\Delta,1-\Delta), &\quad$ \bigl[|k|\ge\sigma_L^{-1}
\bigr]$}\label{eq-hatUasy}
\end{eqnarray}
for some $\zeta\in(0,2)$ and $\Delta\in(0,1)$.
(The assumption $|\hat U_L(k)|<1-\Delta$ is used only to get exponential
decay of $\mathcal{I}_2$ in (\ref{eq-cI2-def}) below.) Combining
these distributions,
we define $D$ as
%
%e4.5 #&#
\begin{equation}
\label{eqeg123-again} D(x)=\sum_{t\in\mathbb{N}}U_L^{*t}(x)
T_\alpha(t).
\end{equation}

We note that the above definition is a discrete version of the transition
kernel for the so-called subordinate process (e.g., \cite{bg68}). Just like
(\ref{eqeg123-again}), the transition kernel for the subordinate
process is
given by an integral of the Gaussian density with respect to the 1-dimensional
$\alpha/2$-stable distribution. Bogdan and Jakubowski \cite{bj07}
make the
most of this integral representation to estimate derivatives of the transition
kernel. This is close to what we want: to prove (\ref{eq-HK2}). However,
in the
current discrete space--time setting, we cannot simply adopt their
proof to
show (\ref{eq-HK2}). To overcome this difficulty, we will approximate
the lattice
distribution $U_L^{*t}$ in (\ref{eqeg123-again}) by a Gaussian density
(multiplied by a polynomial) by using a discrete version of the
Cram\'er--Edgeworth expansion \cite{br10}, Corollary~22.3.

Before doing so, we first show that the above $D$ satisfies (\ref{eq-hatDasy})
and (\ref{eq-Dbd}).

\begin{pf*}{Verification of (\ref{eq-hatDasy}) and (\ref{eq-Dbd})}
Due to the above definition of $U_L$, we can follow the same argument
as in
\cite{hs02}, Appendix~A, to verify the bound on $1-\hat D$ in (\ref{eq-Dbd}).
Moreover, if (\ref{eq-hatDasy}) is also verified, then we can follow
the same
argument as in \cite{csI}, Appendix~A, to confirm the bound on
$\|D^{*n}\|_\infty$ in (\ref{eq-Dbd}) as well.

It remains to verify (\ref{eq-hatDasy}) for small $k$. First, we note that
%
%e4.6 #&#
\begin{equation}
\label{eq-1-hatD} 1-\hat D(k)=\sum_{t\in\mathbb{N}} \bigl(1-\hat
U^t \bigr) T_\alpha(t) =(1-\hat U)\sum
_{t\in\mathbb{N}}T_\alpha(t)\sum_{s=1}^t
\hat U^{s-1},
\end{equation}
where $\hat U$ is an abbreviation for $\hat U_L(k)$. If $\alpha>2$,
we can take any $\xi\in(0,\alpha/2-1)$ to obtain
\begin{eqnarray}
&& 1-\hat D(k)\nonumber
\\[-2pt]
&&\qquad = (1-\hat U)\sum_{t\in\mathbb{N}}T_\alpha(t)
\sum_{s=1}^t1-(1-\hat U) \sum
_{t\in\mathbb{N}}T_\alpha(t)\sum_{s=1}^t
\bigl(1-\hat U^{s-1} \bigr)
\\
&&\qquad =(1-\hat U)\sum_{t\in\mathbb{N}}t T_\alpha(t)+O
\bigl((1-\hat U)^{1+\xi} \bigr),\nonumber
\end{eqnarray}
where we have used the inequality
%
%e4.7 #&#
\begin{eqnarray}
&&\sum_{t\in\mathbb{N}}T_\alpha(t)\sum
_{s=1}^t \bigl(1-\hat U^{s-1} \bigr)\nonumber
\\
&&\qquad =(1-\hat U)^\xi\sum_{t\in\mathbb{N}}T_\alpha(t)
\sum_{s=1}^t \biggl(\frac{1-\hat U^{s-1}} {
1-\hat U}
\biggr)^\xi \bigl(1-\hat U^{s-1} \bigr)^{1-\xi}
\\
&&\qquad \le2^{1-\xi}(1-\hat U)^\xi\sum
_{t\in\mathbb{N}}t^{1+\xi} T_\alpha(t)=O \bigl((1-\hat
U)^\xi \bigr).
\nonumber
\end{eqnarray}
This together with (\ref{eq-sigmaL2})--(\ref{eq-hatUasy}) implies
(\ref{eq-hatDasy}) for
$\alpha>2$, with $\varepsilon=\zeta\wedge(2\xi)$ and
%
%e4.8 #&#
\begin{equation}
\label{eq-valphaalpha>2} v_\alpha=\frac{\sigma_L^2}{2d}\sum
_{t\in\mathbb{N}}t T_\alpha(t)=O \bigl(L^2 \bigr).
\end{equation}

If $\alpha\in(0,2)$, on the other hand, we first rewrite (\ref{eq-1-hatD})
for small $k$ by setting $\hat u\equiv\log1/\hat U$ and changing the order
of summations as
%
%e4.9 #&#
\begin{eqnarray}
\label{eq-1-hatD<2} 1-\hat D(k)&=&\frac{1-\hat U}{\hat U}\sum
_{t\in\mathbb{N}}T_\alpha(t)\sum_{s=1}^te^{-
\hat us}
\nonumber\\[-8pt]\\[-8pt]
&=&\frac{1-\hat U}{\hat U}\frac{\sum_{s\in\mathbb{N}}e^{-\hat
us}\sum_{t=s}^\infty
t^{-1-\alpha/2}}{\sum_{s\in\mathbb{N}}s^{-1-\alpha/2}}.\nonumber
\end{eqnarray}
We note that, for small $k$,
%
%e4.10 #&#
\begin{equation}
\qquad \frac{1-\hat U}{\hat U}=1-\hat U+O \bigl((1-\hat U)^2 \bigr),\qquad\hat
u=1-\hat U+O \bigl((1-\hat U)^2 \bigr).
\end{equation}
Therefore, by a Riemann-sum approximation, we can estimate the
numerator in~(\ref{eq-1-hatD<2}) as
%
%e4.11 #&#
\begin{eqnarray}
\label{eq-1-hatD<2-2} &&\sum_{s\in\mathbb{N}}e^{-\hat us}\sum
_{t=s}^\infty t^{-1-\alpha /2}\nonumber
\\
&&\qquad =\sum_{s\in\hat u\mathbb{N}}e^{-s}\mathop{\sum
_{t\in\hat u\mathbb{N}}}_{(t\ge s)} \biggl(\frac{t}{\hat u}
\biggr)^{-1-\alpha/2}
\nonumber\\[-8pt]\\[-8pt]
&&\qquad =\hat u^{\alpha/2-1} \bigl(1+O(\hat u) \bigr)\int
_0^\infty\mathrm{d}s\,e^{-s} \int
_s^\infty\mathrm{d}t\,t^{-1-\alpha/2}\nonumber
\\
&&\qquad =\frac{2}\alpha\Gamma(1-\alpha/2) \hat u^{\alpha/2-1}
\bigl(1+O( \hat u) \bigr).\nonumber
\end{eqnarray}
This together with (\ref{eq-sigmaL2})--(\ref{eq-hatUasy}) and (\ref{eq-valphaalpha>2})--(\ref{eq-1-hatD<2-2})
implies (\ref{eq-hatDasy}) for $\alpha\in(0,2)$, with $\varepsilon
=\zeta$ and
%
%e4.12 #&#
\begin{equation}
v_\alpha=\frac{2}\alpha\frac{\Gamma(1-\alpha/2)}{\sum_{s\in
\mathbb{N}}s^{-1-\alpha/2}} \biggl(
\frac{\sigma_L^2}{2d} \biggr)^{\alpha/2}=O \bigl(L^\alpha \bigr).
\end{equation}
This verifies that $D$ in (\ref{eqeg123-again}) satisfies both (\ref{eq-hatDasy})
and (\ref{eq-Dbd}).\noqed
\end{pf*}

\begin{pf*}{Verification of (\ref{eq-D-def}), (\ref{eq-HK1}) and
(\ref{eq-HK2})}
To verify these $x$-space bounds on the transition probability $D^{*n}$ and
its discrete derivative, we use the Cram\'er--Edgeworth expansion to
approximate the lattice distribution $U_L^{*t}(x)$ in (\ref{eqeg123-again})
to the Gaussian density $\nu_{\sigma_L^2t}(x)$ (multiplied by a
polynomial of
$x/\sqrt{\sigma_L^2t}$), where
%
%e4.13 #&#
\begin{equation}
\label{eq-normal} \nu_c (x)= \biggl(\frac{d}{2\pi c}
\biggr)^{d/2}\exp \biggl(-\frac{d|x|^2}{2c} \biggr).
\end{equation}

Before showing a precise statement (cf., Theorem~\ref{theorembr10} below),
we explain the formal expansion (\ref{eq-U*t-simp2}) of $U_L^{*t}(x)$. First,
we note that $\hat U_L(k)$ is a generating function of cumulants
$Q_{\vec n}$
for $\vec n\in\mathbb{Z}_+^d$:
%
%e4.14 #&#
\begin{equation}
\label{eq-Qn-def} \log\hat U_L(k)=\mathop{\sum
_{\vec n\in\mathbb{Z}_+^d}}_{(\|\vec
n\|_1\ge1)} Q_{\vec n}\prod
_{s=1}^d\frac{(ik_s)^{n_s}}{n_s!}.
\end{equation}
Since $U_L$ is $\mathbb{Z}^d$-symmetric, we have $Q_{\vec n}=0$ if $\|
\vec n\|
_1$ is
odd, and $Q_{(2,0,\ldots,0)}=\cdots=Q_{(0,\ldots,0,2)}=\sigma_L^2/d$.
Therefore,
%
%e4.15 #&#
\begin{equation}
\label{eq-Qn-rewr} \log\hat U_L(k)=-\frac{\sigma_L^2}{2d}|k|^2+
\sum_{l=4}^\infty\mathop{\sum
_{\vec n\in\mathbb{Z}_+^d}}_{(\|\vec n\|_1=l)} Q_{\vec n}\prod
_{s=1}^d\frac{(ik_s)^{n_s}}{n_s!}.
\end{equation}
By the Fourier inversion theorem, we may rewrite $U_L^{*t}(x)$ as
%
%e4.16 #&#
\begin{eqnarray}\label{eq-U*t-simp1}
U_L^{*t}(x) &=&\int_{[-\pi,\pi]^d}\frac{\mathrm{d}^dk}{(2\pi)^d}\hat U_L(k)^te^{-ik\cdot x}\nonumber
\\
&=&\int_{[-\pi,\pi]^d}\frac{\mathrm{d}^dk}{(2\pi)^d}e^{-(\sigma_L^2/(2d))t
|k|^2-ik\cdot x}\nonumber
\\
&&\hspace*{35pt}{} \times\exp \Biggl(t\sum_{l=4}^\infty \mathop{\sum_{\vec n\in\mathbb{Z}_+^d}}_{(\|\vec n\|_1=l)}Q_{\vec n}
\prod_{s=1}^d \frac{(ik_s)^{n_s}}{n_s!} \Biggr)
\\
&=& \bigl(\sigma_L^2t \bigr)^{-d/2}\int_{\sqrt{\sigma_L^2t}[-\pi,\pi]^d}\frac{\mathrm{d}^dk} {(2\pi)^d}e^{-(1/(2d))|k|^2-ik\cdot\tilde x}\nonumber
\\
&&\hspace*{100pt}{} \times\exp \Biggl(\sum_{l=4}^\infty t^{1-l/2}\tilde Q_l(ik) \Biggr), \nonumber
\end{eqnarray}
where, in the third equality, we have replaced $k$ by $k/\sqrt{\sigma_L^2t}$
and used the abbreviations
%
%e4.17 #&#
\begin{equation}
\label{eq-tildex-def} \tilde x=\frac{x}{\sqrt{\sigma_L^2t}},\qquad \tilde Q_l(ik)=
\mathop{\sum_{\vec n\in\mathbb{Z}_+^d}}_{(\|\vec n\|_1=l)}
\frac{Q_{\vec n}}{\sigma_L^l}\prod_{s=1}^d
\frac{(ik_s)^{n_s}}{n_s!}.
\end{equation}
Notice that, since $U_L$ is supported in $[-L,L]^d$, the coefficients
$Q_{\vec n}/\sigma_L^l$ for \mbox{$\|\vec n\|_1=l$} are uniformly bounded in $L$.
Then the exponential factor involving higher-order cumulants in
(\ref{eq-U*t-simp1}) may be expanded as
%
%e4.18 #&#
\begin{eqnarray}
\label{eq-exp-expansion} &&\exp \Biggl(\sum_{l=2}^\infty
t^{-l/2}\tilde Q_{l+2}(ik) \Biggr)\nonumber
\\
&&\qquad =1+\sum_{m=1}^\infty
\frac{1}{m!} \sum_{l_1,\ldots,l_m\ge2}\,\prod
_{r=1}^m \bigl(t^{-l_r/2}\tilde
Q_{l_r+2}(ik) \bigr)
\\
&&\qquad =1+\sum_{j=2}^\infty
t^{-j/2}\sum_{m=1}^{\lfloor j/2\rfloor}
\frac{1}{m!} \mathop{\sum_{l_1,\ldots,l_m\ge2}}_{(l_1+\cdots+l_m=j)}
\prod_{r=1}^m \tilde Q_{l_r+2}(ik).
\nonumber
\end{eqnarray}
Let
%
%e4.19 #&#
\begin{eqnarray}\label{eq-Pj-def}
P_0(ik) &=& 1,\qquad P_1(ik)=0,
\nonumber\\[2pt]\\[-18pt]
P_j(ik)&=&\sum_{m=1}^{\lfloor j/2\rfloor}
\frac{1}{m!}\mathop{\sum_{l_1,\ldots,l_m\ge2}}_{(l_1+\cdots+l_m=j)}
\prod_{r=1}^m\tilde Q_{l_r+2}(ik)
\qquad[j\ge2].\nonumber
\end{eqnarray}
Then, by (\ref{eq-U*t-simp1}) and (\ref{eq-exp-expansion}), we arrive
at the formal
Cram\'er--Edgeworth expansion
%
%e4.20 #&#
\begin{eqnarray}
\label{eq-U*t-simp2} \qquad U_L^{*t}(x)&=& \bigl(
\sigma_L^2t \bigr)^{-d/2}
\nonumber\\[-2pt]\\[-14pt]
&&{}\times \int_{\sqrt{\sigma_L^2t}[-\pi,\pi]^d} \frac{\mathrm
{d}^dk}{(2\pi)^d}e^{-(1/(2d))|k|^2-ik\cdot\tilde x} \sum
_{j=0}^\infty t^{-j/2}P_j(ik).\nonumber
\end{eqnarray}

Now we note that, if $\sqrt{\sigma_L^2t}[-\pi,\pi]^d$ is replaced by
$\mathbb{R}^d$,
if $\sum_{j=0}^\infty$ is replaced by $\sum_{j=0}^\ell$ for some
$\ell
<\infty$,
and if $x$ is considered to be an element of~$\mathbb{R}^d$ instead of~$\mathbb{Z}^d$,
then we
obtain
%
%e4.21 #&#
\begin{eqnarray}
\label{eq-U*t-simp3} && \bigl(\sigma_L^2t
\bigr)^{-d/2} \int_{\mathbb{R}^d}\frac{\mathrm{d}^dk}{(2\pi
)^d}e^{-(1/((2d))|k|^2-ik\cdot\tilde x}\sum_{j=0}^\ell t^{-j/2}P_j(ik)\nonumber
\\
&&\qquad = \bigl(\sigma_L^2t \bigr)^{-d/2}\sum
_{j=0}^\ell t^{-j/2}\tilde
P_j\int_{\mathbb{R}^d} \frac{\mathrm{d}^dk}{(2\pi)^d}e^{-(1/(2d))|k|^2-ik\cdot\tilde x}
\\
&&\qquad = \bigl(\sigma_L^2t \bigr)^{-d/2}\sum
_{j=0}^\ell t^{-j/2}\tilde
P_j\nu_1(\tilde x),
\nonumber
\end{eqnarray}
where $\tilde P_j$ is the differential operator defined by replacing each
$ik_s$ of $P_j(ik)$ in~(\ref{eq-Pj-def}) by $-\partial/\partial
\tilde x_s$:
%
%e4.22 #&#
\begin{equation}\label{eq-tildePj-def}
\qquad \tilde P_0=1,\qquad\tilde P_1=0,
\qquad\tilde P_j=P_j \biggl(\frac{-\partial}{\partial\tilde x_1},\ldots,\frac
{-\partial} {
\partial\tilde x_d} \biggr)\qquad[j\ge2].
\end{equation}
Notice that, by (\ref{eq-tildex-def}) and (\ref{eq-Pj-def}),
%
%e4.23 #&#
\begin{equation}
\label{eq-interpret} \bigl(\sigma_L^2t
\bigr)^{-d/2} \tilde P_j\nu_1(\tilde x)
=H_{j+2}^{2j} \biggl(\frac{x}{\sqrt{\sigma_L^2t}} \biggr)
\nu_{\sigma
_L^2t}(x),
\end{equation}
where $H_{j+2}^{2j}$ is a polynomial of degree at least $j+2$ and at most
$2j$ (due to the symmetry of $U_L$). The coefficients of the polynomial
are uniformly bounded in $L$, as explained below (\ref{eq-tildex-def}).

The following theorem is a version of \cite{br10}, Corollary~22.3, for
symmetric distributions, which gives a bound on the difference between
$U_L^{*t}(x)$ and (\ref{eq-U*t-simp3}).

%
%th4.1 #&#
\begin{theorem}\label{theorembr10}
For any $x\in\mathbb{Z}^d$, $t\in\mathbb{N}$ and $\ell\in\mathbb{Z}_+$,
%
%e4.24 #&#
\begin{equation}
\label{eq-tildePapprox} \qquad \bigl(1+|\tilde x|^{\ell+2} \bigr) \Biggl|U_L^{*t}(x)-
\bigl(\sigma_L^2t \bigr)^{-d/2}\sum
_{j=0}^\ell t^{-j/2}\tilde
P_j\nu_1(\tilde x) \Biggr|\le\frac
{O(L^{-d})}{t^{(d+\ell)/2}},
\end{equation}
where $\tilde x$ and $\tilde P_j$ are defined in (\ref{eq-tildex-def}) and
(\ref{eq-tildePj-def}), respectively.
\end{theorem}

Before using this theorem to verify (\ref{eq-D-def}),
(\ref{eq-HK1}) and (\ref{eq-HK2}), we briefly explain how to prove
that the
contribution which comes from 1 on the left-hand side\vadjust{\goodbreak} of~(\ref{eq-tildePapprox})
is bounded by $O(L^{-d})t^{-(d+\ell)/2}$, as in (\ref{eq-tildePapprox}).
(To investigate the contribution that comes from $|\tilde x|^{\ell+2}$ on
the left-hand side of (\ref{eq-tildePapprox}), we also use identities
such as
%
%e4.25 #&#
\begin{eqnarray}
\qquad &&\tilde x_1^{\ell+2}U_L^{*t}(x)
\nonumber\\[-2pt]\\[-18pt]
&&\qquad = \bigl(\sigma_L^2t \bigr)^{-d/2}\int
_{\sqrt{\sigma_L^2t}
[-\pi,\pi]^d}\frac{\mathrm{d}^dk}{(2\pi)^d} e^{-ik\cdot\tilde x} \frac
{\partial^{\ell+2}}{\partial(ik_1)^{\ell+2}}
\hat U_L \biggl(\frac{k} {
\sqrt{\sigma_L^2t}} \biggr)^t,
\nonumber
\end{eqnarray}
which is a result of integration by parts.) First, we split the domain of
integration in Fourier space into $E_1=\{k\in\mathbb{R}^d\dvtx|k|\le
\sqrt{t}\}$,
$E_2=\sqrt{\sigma_L^2t}[-\pi,\pi]^d\setminus E_1$ and
$E_3=\mathbb{R}^d\setminus E_1$.
Then the difference between $U_L^{*t}(x)$ and (\ref{eq-U*t-simp3}) is equal
to $\mathcal{I}_1+\mathcal{I}_2-\mathcal{I}_3$, where
%
%e4.26 #&#
%e4.27 #&#
%e4.28 #&#
\begin{eqnarray}
\mathcal{I}_1&=& \bigl(\sigma_L^2t
\bigr)^{-d/2}\int_{E_1}\frac{\mathrm
{d}^dk}{(2\pi)^d} e^{-ik\cdot \tilde x}
\nonumber\\[-8pt]\\[-8pt]
&&\hspace*{60pt}{} \times \Biggl(\hat U_L \biggl(\frac{k}{\sqrt{\sigma
_L^2t}} \biggr)^t-e^{-(1/(2d))|k|^2}\sum_{j=0}^\ell
t^{-j/2}P_j(ik) \Biggr),\hspace*{-35pt}\nonumber
\\
\mathcal{I}_2&=& \bigl(\sigma_L^2t
\bigr)^{-d/2}\int_{E_2}\frac{\mathrm
{d}^dk}{(2\pi)^d}
e^{-ik\cdot
\tilde x} \hat U_L \biggl(\frac{k}{\sqrt{\sigma_L^2t}}
\biggr)^t,\label{eq-cI2-def}
\\
\mathcal{I}_3&=& \bigl(\sigma_L^2t
\bigr)^{-d/2}\int_{E_3}\frac{\mathrm
{d}^dk}{(2\pi)^d}
e^{-(1/(2d))|k|^2-ik\cdot\tilde x}\sum_{j=0}^\ell t^{-j/2}P_j(ik).
\end{eqnarray}
Since (\ref{eq-tildePapprox}) for $t=1$ is trivial, we can assume
$t\ge2$
with no
loss of generality. Then it is not difficult to prove that $\mathcal
{I}_2$ and
$\mathcal{I}_3$
are both bounded by $O(L^{-d})t^{-(d+\ell)/2}$, due to direct
computation for
$\mathcal{I}_3$, and due to (\ref{eq-hatUasy}) and similar
computation to \cite{csI}, (A.2),
for $\mathcal{I}_2$. For $\mathcal{I}_1$, we can bound the integrand by
$Ct^{-\ell/2}(|k|^{\ell+2}+|k|^{2\ell})e^{-c|k|^2}$ for some $L$-independent
constants $C,c\in(0,\infty)$, due to a version of \cite{br10}, Theorem~9.12,
for symmetric distributions. Then, by direct computation,
we can prove that $\mathcal{I}_1$ is also bounded by
$O(L^{-d})t^{-(d+\ell)/2}$.

Now we apply (\ref{eq-tildePapprox}) to verify the $x$-space bounds
(\ref{eq-D-def}),
(\ref{eq-HK1}) and (\ref{eq-HK2}). In particular, by (\ref{eqeg123-again}) and
(\ref{eq-tildePj-def})--(\ref{eq-tildePapprox}),
%
%e4.29 #&#
\begin{eqnarray}
\label{eq-Dest} D(x)&=&\sum_{t=1}^\infty
\nu_{\sigma_L^2t}(x) T_\alpha(t)\nonumber
\\
&&{}+\sum_{t=1}^\infty\sum
_{j=2}^\ell t^{-j/2}H_{j+2}^{2j}
\biggl(\frac{x} {
\sqrt{\sigma_L^2t}} \biggr) \nu_{\sigma_L^2t}(x) T_\alpha(t)
\\
&&{}+\sum_{t=1}^\infty\frac{O(L^{-d})}{t^{(d+\ell)/2}}
\biggl(1\wedge \biggl(\frac{\sqrt{\sigma_L^2t}}{|x|} \biggr)^{\ell+2} \biggr)
T_\alpha(t).
\nonumber
\end{eqnarray}
The leading term is bounded as
%
%e4.30 #&#
\begin{eqnarray}
\label{eq-Dest1}
\qquad \sum_{t=1}^\infty
\nu_{\sigma_L^2t}(x) T_\alpha(t)&\le& O \bigl(L^{-d} \bigr)
\sum_{1 \le t<\veee{x/\sigma_L}_1^2}\frac{\exp(-(d|x|^2)/(2\sigma _L^2t) )} {
t^{1+(d+\alpha)/2}}\nonumber
\\
&&{}+O \bigl(L^{-d} \bigr)\sum_{t\ge\veee{x/\sigma_L}_1^2}t^{-1-(d+\alpha
)/2} \nonumber
\\
&\le& O \bigl(L^{-d} \bigr)\sum_{1\le t<\veee{x/\sigma_L}_1^2}
\frac{O ( ( |x|^2/(\sigma_L^2t) )^{-1-(d+\alpha)/2}
)}{t^{1+(d+\alpha
)/2}}
\\
&&{}+O \bigl(L^{-d} \bigr)\veee{x/\sigma_L}_1^{-(d+\alpha)}
\nonumber
\\
&=& O \bigl(L^\alpha \bigr)\veee{x}_L^{-d-\alpha}.
\nonumber
\end{eqnarray}
The second term on the right-hand side of (\ref{eq-Dest}) is bounded,
due to (\ref{eq-interpret}), as follows: for any $j\in\{2,\ldots,\ell
\}$ and
$h\in\{j+2,\ldots,2j\}$,
%
%e4.31 #&#
\begin{eqnarray}
\label{eq-Dest11} &&\sum_{t=1}^\infty
t^{-j/2} \biggl(\frac{|x|}{\sqrt{\sigma
_L^2t}} \biggr)^h \nu_{\sigma_L^2t}(x) T_\alpha(t)\nonumber
\\
&&\qquad \le O \bigl(L^{-d-h} \bigr)|x|^h\sum _{1\le
t<\veee{x/\sigma_L}_1^2}\frac{\exp(-(d|x|^2)/(2\sigma
_L^2t))} {
t^{1+(d+h+j+\alpha)/2}}
\nonumber\\[-8pt]\\[-8pt]
&&\quad\qquad {} +O \bigl(L^{-d-h} \bigr)|x|^h\sum
_{t\ge\veee{x/\sigma
_L}_1^2}t^{-1-(d+h+j+\alpha)/2}
\nonumber
\\
&&\qquad \le O \bigl(L^{-d-h} \bigr)|x|^h
\frac{O(L^{d+h+j+\alpha})}{\veee
{x}_L^{d+h+j+\alpha}} = \frac{O(L^{j+\alpha})}{\veee{x}_L^{d+j+\alpha}} \le\frac{O(L^{2+\alpha
})}{\veee{x}_L^{d+2+\alpha}}.
\nonumber
\end{eqnarray}
Therefore,
%
%e4.32 #&#
\begin{equation}
\label{eq-Dest2} \sum_{t=1}^\infty\sum
_{j=2}^\ell t^{-j/2}H_{j+2}^{2j}
\biggl(\frac
{x}{\sqrt{\sigma_L^2
t}} \biggr) \nu_{\sigma_L^2t}(x) T_\alpha(t)\le
\frac{O(L^{\alpha+2})} {
\veee{x}_L^{d+\alpha+2}}.
\end{equation}
Similarly, the third term on the right-hand side of (\ref{eq-Dest}) is
bounded as
%
%e4.33 #&#
\begin{eqnarray}
\label{eq-Dest3} &&\sum_{t=1}^\infty
\frac{O(L^{-d})}{t^{(d+\ell)/2}} \biggl(1\wedge \biggl(\frac{\sqrt {\sigma_L^2t}}{|x|}
\biggr)^{\ell+2} \biggr) T_\alpha(t)\nonumber
\\
&&\qquad \le O \bigl(L^{-d+\ell+2} \bigr)|x|^{-\ell-2}\sum
_{1\le t<\veee{x /\sigma_L}_1^2}t^{-(d+\alpha)/2} \nonumber
\\
&&\quad\qquad {} +O \bigl(L^{-d} \bigr)\sum_{t\ge\veee{x/\sigma_L}_1^2}t^{-1-(d+\ell +\alpha)/2}
\nonumber\\[-36pt]\\[20pt]
&&\qquad =\cases{  O \bigl(L^{-d+\ell+2} \bigr)
\veee{x}_L^{-\ell-2}, &\quad$[d+\alpha>2]$,
\vspace*{4pt}\cr
\displaystyle O \bigl(L^{-d+\ell+2} \bigr) \veee{x}_L^{-\ell-2}\log
\veee{x/\sigma_L}_1,& \quad$[d+\alpha=2]$,
\vspace*{4pt}\cr
\displaystyle O \bigl(L^{\alpha+\ell} \bigr) \veee{x}_L^{-d-\alpha-\ell},
&\quad$[d+\alpha<2]$,}\nonumber
\end{eqnarray}
which is further bounded by $O(L^{\alpha+2})\veee{x}_L^{-d-\alpha
-2}$ for
sufficiently large $\ell$. Summarizing the above estimates, we can conclude
(\ref{eq-D-def}):
%
%e4.34 #&#
\begin{equation}
\label{eq-Destfin} D(x)=\sum_{t=1}^\infty
\nu_{\sigma_L^2t}(x) T_\alpha(t)+\frac
{O(L^{\alpha+2})} {
\veee{x}_L^{d+\alpha+2}}\le
\frac{O(L^\alpha)}{\veee
{x}_L^{d+\alpha}}.
\end{equation}

The bound (\ref{eq-HK1}) on the $n$-step transition probability is then
automatically verified, due to the argument below (\ref{eq-HK1}).
Heuristically, since
%
%e4.35 #&#
\begin{eqnarray}
\label{eq-D*n} D^{*n}(x)&=&\sum_{t=n}^\infty
U_L^{*t}(x) T_\alpha^{*n}(t),
\end{eqnarray}
this suggests that
%
%e4.36 #&#
\begin{equation}
\label{eq-T*n} T_\alpha^{*n}(t)\le O(n) T_{\alpha\wedge2}(t).
\end{equation}
In fact, we can verify this (or a stronger version) by following the same
argument as given below (\ref{eq-HK1}), but we omit the details here.

Finally, we verify (\ref{eq-HK2}) by using (\ref{eq-tildePapprox}) with
sufficiently large $\ell$ and (\ref{eq-Destfin})--(\ref{eq-T*n}).
For $|y|\le\frac{1}3|x|$ (so that $|x\pm y|\ge\frac{2}3|x|$), we obtain
%
%e4.37 #&#
\begin{eqnarray}\label{eq-discr-derv}
\quad &&D^{*n}(x)-\frac{D^{*n}(x+y)+D^{*n}(x-y)}2 \nonumber
\\
&&\qquad =\sum_{t=1}^\infty \biggl(\frac
\eta{\pi t} \biggr)^{d/2} \biggl(e^{-\eta|x|^2/t}- \frac{e^{-\eta|x+y|^2/t}+e^{-\eta
|x-y|^2/t}}2
\biggr)T_\alpha^{*n}(t)
\\
&&\quad\qquad {} +\frac{O
(L^{\alpha\wedge2+2})}{\veee{x}_L^{d+\alpha\wedge2+2}} n, \nonumber
\end{eqnarray}
where we have set $\eta=d/(2\sigma_L^2)=O(L^{-2})$ for convenience.
By a Taylor expansion,
%
%e4.38 #&#
\begin{equation}
e^{-\eta|x|^2/t}-\frac{e^{-\eta|x+y|^2/t}+e^{-\eta
|x-y|^2/t}}2=\frac
{O(\eta
|y|^2)}t e^{-\eta|x|^2/t}.
\end{equation}
Using this and (\ref{eq-T*n}) and following the same analysis as in
(\ref{eq-Dest1})--(\ref{eq-Dest11}), we can bound the sum in (\ref{eq-discr-derv}) by
%
%e4.39 #&#
\begin{equation}
\label{eq-discr-main} O \bigl(\eta^{1+d/2} \bigr)|y|^2n\sum
_{t=1}^\infty\frac{e^{-\eta
|x|^2/t}}{t^{2+(d+\alpha
\wedge2)/2}}=
\frac{O(\eta^{-(\alpha\wedge2)/2})|y|^2}{\veee
{x}_{1/\sqrt
\eta}^{d+\alpha\wedge2+2}} n.
\end{equation}
This together with (\ref{eq-discr-derv}) and $\veee{y}_L=|y|\vee L$ yields
(\ref{eq-HK2}).\noqed
\end{pf*}
\end{appendix}

\section*{Acknowledgements}
%The work of the first-named author was supported by the NSC Grant
%99-2115-M-030-004-MY3, and the work of the second-named author was
%supported
%in part by the JSPS Grant-in-Aid for Young Scientists (B) 21740059 and
%in
%part by the JSPS Grant-in-Aid for Scientific Research (C) 24540106.
Akira Sakai is grateful to Remco van der Hofstad for encouraging
conversations, and to Panki Kim, Takashi Kumagai and K\^{o}hei Uchiyama for
pointing him to the relevant literature. We would like to thank the anonymous
referees for many useful suggestions to improve presentation of the manuscript.

%suskaldyti doi

% imsref loaded by linak, 2014-01-13 15:46:42
%

\printaddresses


\begin{thebibliography}{28}
% Style name=ims, version=2.7, label_style=nolabel,
%sorting_style=complex, cfg=None, language=None.
%b1 ###
%b1 #&#
\bibitem{a82}
%
\begin{barticle}[mr]
\bauthor{\bsnm{Aizenman},~\bfnm{Michael}\binits{M.}}
(\byear{1982}).
\btitle{Geometric analysis of $\varphi^{4}$ fields and {I}sing models.
{I}, {II}}.
\bjournal{Comm. Math. Phys.}
\bvolume{86}
\bpages{1--48}.
\bid{issn={0010-3616}, mr={0678000}}
\end{barticle}
%
\bptok{imsref}%
% NOT OUTPUTED:
% issn = 0010-3616
% url = http://projecteuclid.org/getRecord?id=euclid.cmp/1103921614
% number = 1
% coden = CMPHAY
% fjournal = Communications in Mathematical Physics
\endbibitem

%b2 ###
%b2 #&#
\bibitem{af86}
%
\begin{barticle}[mr]
\bauthor{\bsnm{Aizenman},~\bfnm{M.}\binits{M.}} \AND
\bauthor{\bsnm{Fern{\'a}ndez},~\bfnm{R.}\binits{R.}}
(\byear{1986}).
\btitle{On the critical behavior of the magnetization in
high-dimensional {I}sing models}.
\bjournal{J. Stat. Phys.}
\bvolume{44}
\bpages{393--454}.
\bid{doi={10.1007/BF01011304}, issn={0022-4715}, mr={0857063}}
\end{barticle}
%
\bptok{imsref}%
% NOT OUTPUTED:
% issn = 0022-4715
% url = http://dx.doi.org/10.1007/BF01011304
% number = 3-4
% coden = JSTPSB
% fjournal = Journal of Statistical Physics
\endbibitem

%b3 ###
%b3 #&#
\bibitem{an84}
%
\begin{barticle}[mr]
\bauthor{\bsnm{Aizenman},~\bfnm{Michael}\binits{M.}} \AND
\bauthor{\bsnm{Newman},~\bfnm{Charles~M.}\binits{C.~M.}}
(\byear{1984}).
\btitle{Tree graph inequalities and critical behavior in percolation models}.
\bjournal{J. Stat. Phys.}
\bvolume{36}
\bpages{107--143}.
\bid{doi={10.1007/BF01015729}, issn={0022-4715}, mr={0762034}}
\end{barticle}
%
\bptok{imsref}%
% NOT OUTPUTED:
% issn = 0022-4715
% url = http://dx.doi.org/10.1007/BF01015729
% number = 1-2
% coden = JSTPSB
% fjournal = Journal of Statistical Physics
\endbibitem

%b4 ###
%b4 #&#
\bibitem{an86}
%
\begin{barticle}[mr]
\bauthor{\bsnm{Aizenman},~\bfnm{M.}\binits{M.}} \AND
\bauthor{\bsnm{Newman},~\bfnm{C.~M.}\binits{C.~M.}}
(\byear{1986}).
\btitle{Discontinuity of the percolation density in one-dimensional
$1/\vert x- y\vert^2$ percolation models}.
\bjournal{Comm. Math. Phys.}
\bvolume{107}
\bpages{611--647}.
\bid{issn={0010-3616}, mr={0868738}}
\end{barticle}
%
\bptok{imsref}%
% NOT OUTPUTED:
% issn = 0010-3616
% url = http://projecteuclid.org/getRecord?id=euclid.cmp/1104116233
% number = 4
% coden = CMPHAY
% fjournal = Communications in Mathematical Physics
\endbibitem

%b5 ###
%b5 #&#
\bibitem{ba91}
%
\begin{barticle}[mr]
\bauthor{\bsnm{Barsky},~\bfnm{D.~J.}\binits{D.~J.}} \AND
\bauthor{\bsnm{Aizenman},~\bfnm{M.}\binits{M.}}
(\byear{1991}).
\btitle{Percolation critical exponents under the triangle condition}.
\bjournal{Ann. Probab.}
\bvolume{19}
\bpages{1520--1536}.
\bid{issn={0091-1798}, mr={1127713}}
\end{barticle}
%
\bptok{imsref}%
% NOT OUTPUTED:
% issn = 0091-1798
% url =
%%%%%http://links.jstor.org/sici?sici=0091-1798(199110)19:4&lt;1520:PCEUTT&gt;2.0.CO;2-A&origin=MSN
% number = 4
% coden = APBYAE
% fjournal = The Annals of Probability
\endbibitem

%b6 ###
%b6 #&#
\bibitem{br10}
%
\begin{bbook}[auto:STB|2014/01/06|10:16:28]
\bauthor{\bsnm{Bhattacharya},~\bfnm{R.~N.}\binits{R.~N.}} \AND
\bauthor{\bsnm{Rao},~\bfnm{R.~R.}\binits{R.~R.}}
(\byear{2010}).
\btitle{Normal Approximation and Asymptotic Expansions}.
\bseries{Classics in Applied Mathematics}
\bvolume{64}.
\bpublisher{SIAM}, \blocation{Philadelphia, PA}.
\end{bbook}
%
\bptok{imsref}%
% NOT OUTPUTED:
% sortkey = Bhattacharya(2010
\endbibitem

%b7 ###
%b7 #&#
\bibitem{bg68}
%
\begin{bbook}[mr]
\bauthor{\bsnm{Blumenthal},~\bfnm{R.~M.}\binits{R.~M.}} \AND
\bauthor{\bsnm{Getoor},~\bfnm{R.~K.}\binits{R.~K.}}
(\byear{1968}).
\btitle{Markov Processes and Potential Theory}.
\bpublisher{Academic Press},
\blocation{New York}.
\bid{mr={0264757}}
\end{bbook}
%
\bptok{imsref}%
% NOT OUTPUTED:
% fpage = x+313
\endbibitem

%b8 ###
%b8 #&#
\bibitem{bj07}
%
\begin{barticle}[mr]
\bauthor{\bsnm{Bogdan},~\bfnm{Krzysztof}\binits{K.}} \AND
\bauthor{\bsnm{Jakubowski},~\bfnm{Tomasz}\binits{T.}}
(\byear{2007}).
\btitle{Estimates of heat kernel of fractional {L}aplacian perturbed by
gradient operators}.
\bjournal{Comm. Math. Phys.}
\bvolume{271}
\bpages{179--198}.
\bid{doi={10.1007/s00220-006-0178-y}, issn={0010-3616}, mr={2283957}}
\end{barticle}
%
\bptok{imsref}%
% NOT OUTPUTED:
% issn = 0010-3616
% url = http://dx.doi.org/10.1007/s00220-006-0178-y
% number = 1
% coden = CMPHAY
% fjournal = Communications in Mathematical Physics
\endbibitem

%b9 ###
%b9 #&#
\bibitem{csI}
%
\begin{barticle}[mr]
\bauthor{\bsnm{Chen},~\bfnm{Lung-Chi}\binits{L.-C.}} \AND
\bauthor{\bsnm{Sakai},~\bfnm{Akira}\binits{A.}}
(\byear{2008}).
\btitle{Critical behavior and the limit distribution for long-range
oriented percolation. {I}}.
\bjournal{Probab. Theory Related Fields}
\bvolume{142}
\bpages{151--188}.
\bid{doi={10.1007/s00440-007-0101-2}, issn={0178-8051}, mr={2413269}}
\end{barticle}
%
\bptok{imsref}%
% NOT OUTPUTED:
% issn = 0178-8051
% url = http://dx.doi.org/10.1007/s00440-007-0101-2
% number = 1-2
% coden = PTRFEU
% fjournal = Probability Theory and Related Fields
\endbibitem

%b10 ###
%b10 #&#
\bibitem{csII}
%
\begin{barticle}[mr]
\bauthor{\bsnm{Chen},~\bfnm{Lung-Chi}\binits{L.-C.}} \AND
\bauthor{\bsnm{Sakai},~\bfnm{Akira}\binits{A.}}
(\byear{2009}).
\btitle{Critical behavior and the limit distribution for long-range
oriented percolation. {II}. {S}patial correlation}.
\bjournal{Probab. Theory Related Fields}
\bvolume{145}
\bpages{435--458}.
\bid{doi={10.1007/s00440-008-0174-6}, issn={0178-8051}, mr={2529436}}
\end{barticle}
%
\bptok{imsref}%
% NOT OUTPUTED:
% issn = 0178-8051
% url = http://dx.doi.org/10.1007/s00440-008-0174-6
% number = 3-4
% coden = PTRFEU
% fjournal = Probability Theory and Related Fields
\endbibitem

%b11 ###
%b11 #&#
\bibitem{csIII}
%
\begin{barticle}[mr]
\bauthor{\bsnm{Chen},~\bfnm{Lung-Chi}\binits{L.-C.}} \AND
\bauthor{\bsnm{Sakai},~\bfnm{Akira}\binits{A.}}
(\byear{2011}).
\btitle{Asymptotic behavior of the gyration radius for long-range
self-avoiding walk and long-range oriented percolation}.
\bjournal{Ann. Probab.}
\bvolume{39}
\bpages{507--548}.
\bid{doi={10.1214/10-AOP557}, issn={0091-1798}, mr={2789505}}
\end{barticle}
%
\bptok{imsref}%
% NOT OUTPUTED:
% issn = 0091-1798
% url = http://dx.doi.org/10.1214/10-AOP557
% number = 2
% coden = APBYAE
% fjournal = The Annals of Probability
\endbibitem

%b12 ###
%b12 #&#
\bibitem{g70}
%
\begin{barticle}[mr]
\bauthor{\bsnm{Ginibre},~\bfnm{J.}\binits{J.}}
(\byear{1970}).
\btitle{General formulation of {G}riffiths' inequalities}.
\bjournal{Comm. Math. Phys.}
\bvolume{16}
\bpages{310--328}.
\bid{issn={0010-3616}, mr={0269252}}
\end{barticle}
%
\bptok{imsref}%
% NOT OUTPUTED:
% issn = 0010-3616
% fjournal = Communications in Mathematical Physics
\endbibitem

%b13 ###
%b13 #&#
\bibitem{ghs70}
%
\begin{barticle}[mr]
\bauthor{\bsnm{Griffiths},~\bfnm{Robert~B.}\binits{R.~B.}},
\bauthor{\bsnm{Hurst},~\bfnm{C.~A.}\binits{C.~A.}} \AND
\bauthor{\bsnm{Sherman},~\bfnm{S.}\binits{S.}}
(\byear{1970}).
\btitle{Concavity of magnetization of an {I}sing ferromagnet in a
positive external field}.
\bjournal{J. Math. Phys.}
\bvolume{11}
\bpages{790--795}.
\bid{issn={0022-2488}, mr={0266507}}
\end{barticle}
%
\bptok{imsref}%
% NOT OUTPUTED:
% issn = 0022-2488
% fjournal = Journal of Mathematical Physics
\endbibitem

%b14 ###
%b14 #&#
\bibitem{g99}
%
\begin{bbook}[mr]
\bauthor{\bsnm{Grimmett},~\bfnm{Geoffrey}\binits{G.}}
(\byear{1999}).
\btitle{Percolation},
\bedition{2nd} ed.
%Principles of Mathematical Sciences]}
\bpublisher{Springer},
\blocation{Berlin}.
\bid{mr={1707339}}
\end{bbook}
%
\bptok{imsref}%
% NOT OUTPUTED:
% isbn = 3-540-64902-6
% fpage = xiv+444
\endbibitem

%b15 ###
%b15 #&#
\bibitem{h08}
%
\begin{barticle}[mr]
\bauthor{\bsnm{Hara},~\bfnm{Takashi}\binits{T.}}
(\byear{2008}).
\btitle{Decay of correlations in nearest-neighbor self-avoiding walk,
percolation, lattice trees and animals}.
\bjournal{Ann. Probab.}
\bvolume{36}
\bpages{530--593}.
\bid{doi={10.1214/009117907000000231}, issn={0091-1798}, mr={2393990}}
\end{barticle}
%
\bptok{imsref}%
% NOT OUTPUTED:
% issn = 0091-1798
% url = http://dx.doi.org/10.1214/009117907000000231
% number = 2
% coden = APBYAE
% fjournal = The Annals of Probability
\endbibitem

%b16 ###
%b16 #&#
\bibitem{hhs??}
%
\begin{bmisc}[auto:STB|2014/01/06|10:16:28]
\bauthor{\bsnm{Hara},~\bfnm{T.}\binits{T.}},
\bauthor{\bsnm{Heydenreich},~\bfnm{M.}\binits{M.}} \AND
\bauthor{\bsnm{Sakai},~\bfnm{A.}\binits{A.}}
\bhowpublished{One-arm exponent for the Ising ferromagnets in high
dimensions. In preparation.}
\end{bmisc}
%
\bptok{imsref}%
% NOT OUTPUTED:
% sortkey = Hara
\endbibitem

%b17 ###
%b17 #&#
\bibitem{hs90}
%
\begin{barticle}[mr]
\bauthor{\bsnm{Hara},~\bfnm{Takashi}\binits{T.}} \AND
\bauthor{\bsnm{Slade},~\bfnm{Gordon}\binits{G.}}
(\byear{1990}).
\btitle{Mean-field critical behaviour for percolation in high dimensions}.
\bjournal{Comm. Math. Phys.}
\bvolume{128}
\bpages{333--391}.
\bid{issn={0010-3616}, mr={1043524}}
\end{barticle}
%
\bptok{imsref}%
% NOT OUTPUTED:
% issn = 0010-3616
% url = http://projecteuclid.org/getRecord?id=euclid.cmp/1104180434
% number = 2
% coden = CMPHAY
% fjournal = Communications in Mathematical Physics
\endbibitem

%b18 ###
%b18 #&#
\bibitem{hhs03}
%
\begin{barticle}[mr]
\bauthor{\bsnm{Hara},~\bfnm{Takashi}\binits{T.}},
\bauthor{\bparticle{van~der}~\bsnm{Hofstad},~\bfnm{Remco}\binits{R.}}
\AND
\bauthor{\bsnm{Slade},~\bfnm{Gordon}\binits{G.}}
(\byear{2003}).
\btitle{Critical two-point functions and the lace expansion for
spread-out high-dimensional percolation and related models}.
\bjournal{Ann. Probab.}
\bvolume{31}
\bpages{349--408}.
\bid{doi={10.1214/aop/1046294314}, issn={0091-1798}, mr={1959796}}
\end{barticle}
%
\bptok{imsref}%
% NOT OUTPUTED:
% issn = 0091-1798
% url = http://dx.doi.org/10.1214/aop/1046294314
% number = 1
% coden = APBYAE
% fjournal = The Annals of Probability
\endbibitem

%b19 ###
%b19 #&#
\bibitem{hhh11}
%
\begin{bmisc}[auto]
\bauthor{\bsnm{Heydenreich},~\bfnm{Markus}\binits{M.}},
\bauthor{\bparticle{van~der}~\bsnm{Hofstad},~\bfnm{Remco}\binits{R.}}
\AND
\bauthor{\bsnm{Hulshof},~\bfnm{T.}\binits{T.}}
(\byear{2011}).
\bhowpublished{High-dimensional incipient infinite clusters revisited.
Preprint. Available at \arxivurl{arXiv:1108.4325}.}
\end{bmisc}
%
\bptok{imsref}%
% NOT OUTPUTED:
% issn = 0178-8051
% url = http://dx.doi.org/10.1007/s00440-009-0258-y
% number = 3-4
% coden = PTRFEU
% fjournal = Probability Theory and Related Fields
\endbibitem

%b20 ###
%b20 #&#
\bibitem{hhs08}
%
\begin{barticle}[mr]
\bauthor{\bsnm{Heydenreich},~\bfnm{Markus}\binits{M.}},
\bauthor{\bparticle{van~der}~\bsnm{Hofstad},~\bfnm{Remco}\binits{R.}}
\AND
\bauthor{\bsnm{Sakai},~\bfnm{Akira}\binits{A.}}
(\byear{2008}).
\btitle{Mean-field behavior for long- and finite range {I}sing model,
percolation and self-avoiding walk}.
\bjournal{J. Stat. Phys.}
\bvolume{132}
\bpages{1001--1049}.
\bid{doi={10.1007/s10955-008-9580-5}, issn={0022-4715}, mr={2430773}}
\end{barticle}
%
\bptok{imsref}%
% NOT OUTPUTED:
% issn = 0022-4715
% url = http://dx.doi.org/10.1007/s10955-008-9580-5
% number = 6
% fjournal = Journal of Statistical Physics
\endbibitem

%b21 ###
%b21 #&#
\bibitem{kn11}
%
\begin{barticle}[mr]
\bauthor{\bsnm{Kozma},~\bfnm{Gady}\binits{G.}} \AND
\bauthor{\bsnm{Nachmias},~\bfnm{Asaf}\binits{A.}}
(\byear{2011}).
\btitle{Arm exponents in high dimensional percolation}.
\bjournal{J.~Amer. Math. Soc.}
\bvolume{24}
\bpages{375--409}.
\bid{doi={10.1090/S0894-0347-2010-00684-4}, issn={0894-0347}, mr={2748397}}
\end{barticle}
%
\bptok{imsref}%
% NOT OUTPUTED:
% issn = 0894-0347
% url = http://dx.doi.org/10.1090/S0894-0347-2010-00684-4
% number = 2
% fjournal = Journal of the American Mathematical Society
\endbibitem

%b22 ###
%b22 #&#
\bibitem{ms93}
%
\begin{bbook}[mr]
\bauthor{\bsnm{Madras},~\bfnm{Neal}\binits{N.}} \AND
\bauthor{\bsnm{Slade},~\bfnm{Gordon}\binits{G.}}
(\byear{1993}).
\btitle{The Self-Avoiding Walk}.
\bpublisher{Birkh\"auser},
\blocation{Boston, MA}.
\bid{mr={1197356}}
\end{bbook}
%
\bptok{imsref}%
% NOT OUTPUTED:
% isbn = 0-8176-3589-0
% fpage = xiv+425
\endbibitem

%b23 ###
%b23 #&#
\bibitem{s04}
%
\begin{barticle}[mr]
\bauthor{\bsnm{Sakai},~\bfnm{Akira}\binits{A.}}
(\byear{2004}).
\btitle{Mean-field behavior for the survival probability and the
percolation point-to-surface connectivity}.
\bjournal{J. Stat. Phys.}
\bvolume{117}
\bpages{111--130}.
\bid{doi={10.1023/B:JOSS.0000044061.83860.62}, issn={0022-4715}, mr={2098561}}
\end{barticle}
%
\bptok{imsref}%
% NOT OUTPUTED:
% issn = 0022-4715
% url = http://dx.doi.org/10.1023/B:JOSS.0000044061.83860.62
% number = 1-2
% coden = JSTPSB
% fjournal = Journal of Statistical Physics
\endbibitem

%b24 ###
%b24 #&#
\bibitem{s07}
%
\begin{barticle}[mr]
\bauthor{\bsnm{Sakai},~\bfnm{Akira}\binits{A.}}
(\byear{2007}).
\btitle{Lace expansion for the {I}sing model}.
\bjournal{Comm. Math. Phys.}
\bvolume{272}
\bpages{283--344}.
\bid{doi={10.1007/s00220-007-0227-1}, issn={0010-3616}, mr={2300246}}
\end{barticle}
%
\bptok{imsref}%
% NOT OUTPUTED:
% issn = 0010-3616
% url = http://dx.doi.org/10.1007/s00220-007-0227-1
% number = 2
% coden = CMPHAY
% fjournal = Communications in Mathematical Physics
\endbibitem

%b25 ###
%b25 #&#
\bibitem{s06}
%
\begin{bbook}[mr]
\bauthor{\bsnm{Slade},~\bfnm{G.}\binits{G.}}
(\byear{2006}).
\btitle{The Lace Expansion and Its Applications}.
\bseries{Lecture Notes in Math.}
\bvolume{1879}.
\bpublisher{Springer},
\blocation{Berlin}.
%in Saint-Flour, July 6--24, 2004, Edited and with a foreword by Jean
%Picard}.
\bid{mr={2239599}}
\end{bbook}
%
\bptok{imsref}%
% NOT OUTPUTED:
% isbn = 978-3-540-31189-8; 3-540-31189-0
% fpage = xiv+228
\endbibitem

%%b26 ###
%%b26 #&#
%%
%(\byear{2007}).
%probabilities for {M}arkov additive processes}.
%%
%% NOT OUTPUTED:
%% issn = 1083-6489
%% url = http://dx.doi.org/10.1214/EJP.v12-396
%% fjournal = Electronic Journal of Probability

%b27 ###
%b27 #&#
\bibitem{bk85}
%
\begin{barticle}[mr]
\bauthor{\bparticle{van~den}~\bsnm{Berg},~\bfnm{J.}\binits{J.}}
\AND
\bauthor{\bsnm{Kesten},~\bfnm{H.}\binits{H.}}
(\byear{1985}).
\btitle{Inequalities with applications to percolation and reliability}.
\bjournal{J. Appl. Probab.}
\bvolume{22}
\bpages{556--569}.
\bid{issn={0021-9002}, mr={0799280}}
\end{barticle}
%
\bptok{imsref}%
% NOT OUTPUTED:
% issn = 0021-9002
% number = 3
% coden = JPRBAM
% fjournal = Journal of Applied Probability
\endbibitem

%b28 ###
%b28 #&#
\bibitem{hs02}
%
\begin{barticle}[mr]
\bauthor{\bparticle{van~der}~\bsnm{Hofstad},~\bfnm{Remco}\binits{R.}}
\AND
\bauthor{\bsnm{Slade},~\bfnm{Gordon}\binits{G.}}
(\byear{2002}).
\btitle{A generalised inductive approach to the lace expansion}.
\bjournal{Probab. Theory Related Fields}
\bvolume{122}
\bpages{389--430}.
\bid{doi={10.1007/s004400100175}, issn={0178-8051}, mr={1892852}}
\end{barticle}
%
\bptok{imsref}%
% NOT OUTPUTED:
% issn = 0178-8051
% url = http://dx.doi.org/10.1007/s004400100175
% number = 3
% coden = PTRFEU
% fjournal = Probability Theory and Related Fields
\endbibitem

\end{thebibliography}
\end{document}